\documentclass[AMA,LATO1COL]{WileyNJD-v2}

\usepackage{xcolor}
\usepackage{subfigure}
\usepackage{amssymb}
\usepackage{tikz}
\usetikzlibrary{calc}
 
\newcommand{\enma}[1]   {\ensuremath{#1}}

\newcommand{\beq}{\begin{equation}}
\newcommand{\eeq}{\end{equation}}
\newcommand{\bseq}{\begin{subequations}}
\newcommand{\eseq}{\end{subequations}}
\newcommand{\beqn}{\begin{eqnarray}}
\newcommand{\eeqn}{\end{eqnarray}}
\newcommand{\ba}{\begin{array}}
\newcommand{\ea}{\end{array}}
\newcommand{\bct}{\begin{center}}
\newcommand{\ect}{\end{center}}
\newcommand{\btmz}{\begin{itemize}}
\newcommand{\etmz}{\end{itemize}}
\newcommand{\benum}{\begin{enumerate}}
\newcommand{\eenum}{\end{enumerate}}

\newcommand{\norm}[1]{\| #1 \|}                 

\newcommand{\trace}     {\enma{\mathrm{trace}}}

\newcommand{\bv}{{\bf v}}

\newcommand{\del}{\Delta}

\newcommand{\matbegin}{
        \left[
}
\newcommand{\matend}{
        \right]
}

\newcommand{\tbt}[4]{
  \matbegin \begin{array}{cc}
       #1 & #2 \\ #3 & #4
       \end{array} \matend }

\newcommand{\be}{\begin{equation}}
\newcommand{\ee}{\end{equation}}

\newcommand{\cplxs}{ C\kern -.35em \rule{0.03 em}{.7 ex}~   }

\def\complex{\hbox{C\kern -.45em \rule{0.03 em}{1.5 ex}}~}

\newcommand{\bi}{\begin{itemize}}
\newcommand{\ei}{\end{itemize}}

\newcommand{\bu}{{\bf u}}

\newcommand{\bw}{{\bf w}}

\newcommand{\bbR}{\mathbb{R}}

\newcommand{\btab}{\begin{tabular}}
\newcommand{\etab}{\end{tabular}}
\newcommand{\bd}{{\bf d}}

\newcommand{\bx}{{\bf x}}

\newcommand{\bpsi}{\mbox{\boldmath$\psi$}}

\newcommand{\non}{\nonumber}
\newcommand{\mrd}{\mathrm{d}}

\newcommand{\ds}{\displaystyle}

\newcommand{\bA}{\mathbf{A}}
\newcommand{\bB}{\mathbf{B}}
\newcommand{\bC}{\mathbf{C}}

\newcommand{\bE}{\mathbf{E}}

\newcommand{\DefinedAs}[0]{\mathrel{\mathop:}=}
\newcommand{\vsp}{\vspace*{0.15cm}}

\DeclareMathOperator*{\logdet}{log\,det}
\DeclareMathOperator*{\minimize}{minimize}
\DeclareMathOperator*{\subject}{subject~to}

\definecolor{bgblue}{rgb}{0.04,0.19,0.53}
\definecolor{dblue1}{rgb}{0,0.3,0.7}
\definecolor{dred}{rgb}{0.4,0.2,0}

\DeclareMathOperator{\rank}{rank}

\definecolor{dred}{rgb}{0.4,0.2,0}
\definecolor{orange}{rgb}{1,0.5,0}
\definecolor{bblue}{RGB}{0,0,255}
\definecolor{bgblue}{rgb}{0.04,0.39,0.53}
\definecolor{orange}{rgb}{.232,.117,0.0}
\definecolor{bgblue1}{rgb}{0.05,0.30,0.43}
\definecolor{DarkGreen}{rgb}{.26,.84,.63}   
\definecolor{grn}{rgb}{0.1,0.65,0.1}   
\definecolor{white}{rgb}{1,1,1}
\definecolor{black}{rgb}{0,0,0}
\definecolor{backgrey}{rgb}{0.86,0.86,0.86}
\definecolor{dblue}{rgb}{0,0,0.5}
\definecolor{dblue1}{rgb}{0,0.3,0.7}
\definecolor{dred}{rgb}{0.4,0.2,0}
\definecolor{blue1}{rgb}{0.1, 0.6, 1.0}
\definecolor{blue(ryb)}{rgb}{0.01, 0.28, 1.0}
\definecolor{brandeisblue}{rgb}{0.0, 0.44, 1.0}
\definecolor{silver}{cmyk}{0,0,0,0.3}
\definecolor{yellow}{cmyk}{0,0,0.9,0.0}
\definecolor{reddishyellow}{cmyk}{0,0.22,1.0,0.0}
\definecolor{black}{cmyk}{0,0,0.0,1.0}
\definecolor{darkYellow}{cmyk}{0,0,1.0,0.5}
\definecolor{darkSilver}{cmyk}{0,0,0,0.1}
\definecolor{lightyellow}{cmyk}{0,0,0.3,0.0}
\definecolor{lighteryellow}{cmyk}{0,0,0.1,0.0}
\definecolor{lighteryellow}{cmyk}{0,0,0.1,0.0}
\definecolor{lightestyellow}{cmyk}{0,0,0.05,0.0}
\definecolor{orange}{rgb}{1,0.5,0}
\definecolor{bblue}{RGB}{0,191,205}
\definecolor{lightblue}{rgb}{0.145,0.6666,1}

\articletype{Research Article}%

\received{}
\revised{}
\accepted{}

\begin{document}

\title{Stochastic dynamical wake modeling for wind farms}

\author[]{Aditya H. Bhatt}

\author[]{Federico Bernardoni}

\author[]{Stefano Leonardi}

\author[]{Armin Zare*}

\authormark{Bhatt \textsc{et al.}}

\address[]{\orgdiv{Department of Mechanical Engineering}, \orgname{University of Texas at Dallas}, \orgaddress{\city{Richardson}, \state{Texas}, \country{USA}}}




\corres{*Armin Zare, Department of Mechanical Engineering,
University of Texas at Dallas,
Richardson, TX 75080, USA \\ \email{armin.zare@utdallas.edu}}


\abstract[Abstract]{Low-fidelity analytical models of turbine wakes have traditionally been used for wind farm planning, performance evaluation, and demonstrating the utility of advanced control algorithms in increasing the annual energy production. In practice, however, it remains challenging to correctly estimate the flow and achieve significant performance gains using conventional low-fidelity models. This is due to the over-simplified static nature of wake predictions from models that are agnostic to the effects of atmospheric boundary layer turbulence and the complex aerodynamic interactions among wind turbines. To improve the predictive capability of low-fidelity models while remaining amenable to control design, we offer a stochastic dynamical modeling framework for capturing the effect of atmospheric turbulence on the thrust force and power generation as determined by the actuator disk concept. In this approach, we use stochastically forced linear models of the turbulent velocity field to augment the analytically computed wake velocity and achieve consistency with higher-fidelity models in capturing power and thrust force measurements. The power-spectral densities of our stochastic models are identified via convex optimization to ensure consistency with partially available velocity statistics or power and thrust force measurements while preserving model parsimony. We demonstrate the utility of our approach in capturing turbulence intensity variations behind wind turbines and estimating thrust force and power signals generated by large-eddy simulations of the flow over a cascade of turbines. Our results provide insight into the significance of sparse field measurements in recovering the statistical signature of wind farm turbulence using stochastic linear models.}

\keywords{Convex optimization, data-enhanced control-oriented modeling, stochastically forced Navier-Stokes equations, turbulence modeling, wake modeling, wind energy}


\maketitle

\section{Introduction}
\label{sec.intro}

In recent years, experiment- and simulation-based studies have demonstrated the efficacy of induction and wake steering as control strategies that can improve the performance of wind farms~\cite{fleannshawananazhahutwancheche17,ahmbasahscougirkazmat19,duccougirgiegoc19,howleldab19,flekinsimroaschmurlun20,bosrui20,doekermatkan21,simflegirall21}. With the exception of a small number of studies that have pursued model-free methods for maximizing power production, e.g., extremum seeking control~\cite{johfri12,crelisee09,cirrotleo17,cirrotsanleo17,cirleorot19,kumrot22}, efforts in designing wind farm controllers have relied on various levels of abstraction offered by models of wind farm flows. To date, most model-based approaches have focused on open-loop control policies informed by look-up tables that determine the optimal turbine settings offline and based on the response of static engineering models to different steady-state atmospheric conditions (e.g., wind directions, wind speed, turbulence intensity, etc.). While successful in controlled experimental or numerical testing environments~\cite{howleldab19,bosrui20,flekinsimroaschmurlun20}, potentially unforeseen variations in turbulent inflow conditions, terrain specific effects, or sensing/actuation errors can hinder the generalizability, and thus, applicability of open-loop strategies at the scale of large wind farms. Robust feedback control provides the systematic means to tackle such challenges by systematically accounting for uncertainties in sensing and actuation, unknown exogenous disturbances, and modelling errors~\cite{zhodoy98,skopos07}.

\vsp
Due to the vast range of spatio-temporal scales over which coherent structures affect turbine performance and the intricate nature of wake turbulence, model-based closed-loop control design has predominantly relied on computationally expensive high-fidelity models such as those that are used in large-eddy simulations (LES) to demonstrate meaningful performance improvements~\cite{goimey15,goimunmey16,munmey17,munmey18,munmey18ENG,doevanfle19,sinsei19,doevamvan20}. While such models play an important role in improving our understanding of wake turbulence, they are not suitable for the development of real-time model-based control strategies that can adapt to time-varying atmospheric conditions informed by supervisory control and data acquisition (SCADA) measurements. This motivates the development of lower fidelity models that capture the essential flow features and quantities of interest for analysis or control.

\vspace*{-1.5ex}
\subsection{Control-oriented wake modeling}
\vspace*{-.5ex}

Seminal efforts toward developing low-fidelity models of turbine wakes focused on two-dimensional (2D) heuristic based methods that capture the reduction in the mean streamwise velocity at hub height for given steady atmospheric conditions~\cite{jen83,katic86,ain88}. Enabled by structural approximations of turbine rotors (e.g., the actuator disk model [ADM]~\cite{burjenshabos11}) more sophisticated variants that observe conservation principles for mass and momentum~\cite{frandsen06} or even model the mean streamwise velocity deficit as a Gaussian distribution~\cite{baspor14} were combined with wake superposition laws to improve predictions of the power captured by wind farms. The predictions of such static engineering models of the averaged velocity field typically depend on a set of parameters that can be tuned to match field measurements or LES data (e.g.,~\cite{wantanchogu16,cammolschbot19,zhaletiun20}). Efforts have also been made to incorporate 3D effects resulting from turbine yawing or the ground into numerical integration schemes and predict the curled shape of turbine wakes~\cite{marannflechu19,marbra20,zonpor20}. More recent analytical developments bypass the need for numerical integration while accounting for curled shape deformations and even lateral and wall-normal deflections of turbine wakes (e.g., due to ground effects~\cite{basshashagaymen22}). Nevertheless, in the absence of a dynamical model for the fluctuating velocity field, the over-simplified static nature of conventional engineering wake models that neglects the time-varying features of near-field turbulence leads to the under-prediction of wake recovery. This, in turn, can yield inaccurate predictions of quantities of interest for wind farm control (e.g., the load and power corresponding to each turbine).

\vsp
To overcome the shortcomings of static engineering models, contributions have been made to add a degree of dynamics or parametric stochasticity to analytical models (e.g., the dynamic wake-meandering model~\cite{larmadtholar08}, the dynamic extension of the Park model~\cite{annhowseigua16}, and the stochastic ADM model~\cite{guorotsum20}). Reliance on extensive parametric tuning, dynamical complexities, and the absence of constructive methods for uncertainty modeling challenge the utility of such models for real-time estimation and control. Medium-fidelity models (e.g., those based on the Reynolds-averaged NS [RANS] equations) have sought to overcome these issues by capturing the 3D dynamic variation of the velocity field and incorporating turbulence models~\cite{iunviocirleorot16,boedoevalmeyvan18,letiun22}. However, the limitations of turbulence models and the nonlinearity of RANS-based models hinder their utility for real-time optimal estimation and control.

\vsp
Efforts have also been made to train reduced-order models of low-fidelity using data from numerical simulations. In conjunction with graph-based methods, such data-driven models have been used to estimate the direction of free-stream velocity~\cite{annbayjohdalquokemfle19}, identify clusters of wind turbines within farms~\cite{bercirrotleo21}, or even predict variations in power output due to changes in inlet wind direction~\cite{stastamengaykin21}. In a similar vein, machine learning approaches have been used to obtain reduced-order models based on data collected from experiments and numerical simulations~\cite{vermen14,anngebsei16,raaschche17,sinpaojen20}. Data-driven methods are attractive due to their flexibility in analyzing different physical phenomena. However, unreliable measurements and data anomalies challenge a sole reliance on data because such models are agnostic to the underlying physics. Furthermore, control actuation and sensing may significantly alter the identified modes that are used to construct data-driven models in unpredictable ways. This compromises the performance of the resulting models in regimes that were not accounted for in the training process and introduces nontrivial challenges for model-based control design~\cite{noamortad11,tadnoa11}. The described lack of robustness and generalizability is exacerbated by the uninterpretability of dynamic links that are identified through optimization procedures, the intricate multi-layer nature of models that are identified via (deep) neural networks, and the reliance on extensive parametric tuning. A promising direction is to constrain data-driven models to subspaces that are dictated by the underlying physics, i.e., physics-informed machine learning~\cite{wanwuxia17,wuxiapat18,karkevluperrwanyan21}. An alternative approach, which we pursue in this work, is to leverage the underlying physics in the form of a prior model that arises from first principles, i.e., linearization of the NS equations around stable flow states, and to use data-driven techniques to enhance its predictive capability. 

\vsp
The linearized NS equations have been combined with vortex cylinder theory to provide a physics-based alternative for dynamical modeling of wind farm flows~\cite{solwisbra14}. Furthermore, in conjunction with actuator disk theory, the 2D linearized NS equations have been reformulated as a quasi linear parameter varying descriptor model and used for the purpose of wind farm control~\cite{boevalkuhvan16}. Such models that are based on the linearized NS equations can overcome some of the shortcomings of conventionally used low-fidelity wake models in qualitatively predicting flow features of turbulent wakes and the resulting power production~\cite{boedoevalmeyvan18}. However, quantifying and modeling the uncertainties due to: (i) the choice of base flow around which linearization happens; and (ii) the absence of nonlinear terms, remain challenging. In particular, modeling such sources of uncertainty plays an important role in obtaining well-posed estimation gains when using linear models for predicting flow statistics~\cite{hoechebewhen05,chehoebewhen06} or quantities of interest for control design, e.g., thrust force or power generation at turbines.

\subsection{Stochastic dynamical modeling of turbulent flows}

When subjected to additive stochastic excitation, a linear system provides a stochastic response with statistical characteristics that can be qualitatively and quantitatively compared with the results of nonlinear simulations or experiments. For example, the stochastically forced linearized NS equations have been shown to reproduce structural and statistical features of transitional~\cite{butfar92,tretrereddri93,farioa93,bamdah01,jovbamJFM05,ranzarhacjovPRF19b} and turbulent~\cite{mcksha10,hwacosJFM10b,zarjovgeoJFM17} wall-bounded shear flows. Stochastic excitation captures the effects of neglected nonlinear terms and exogenous excitations. Based on this, inverse problems can be posed to utilize statistical signatures of complex dynamical systems that are generated by numerical simulations (e.g., see~\cite{deljim03,deljimzanmos04,siljimmos13,siljimmos14}) or experimental measurements (e.g., see~\cite{hutmonganngmar11,schfla13}) to shape the statistics of stochastic forcing into the linearized dynamics. This idea was pursued by Zare et al.~\cite{zarchejovgeoTAC17,zarjovgeoJFM17}, where the data-enhanced physics-based modeling framework was developed to account for partially observed statistical signatures of complex dynamical systems (e.g., turbulent flows) by introducing colored-in-time stochastic forcing into the linearized dynamics. The colored-in-time forcing is realized via a systematic optimization-based procedure that ensures the statistical consistency of the output of the linearized dynamics with the result of nonlinear simulations while preserving the principle of parsimony. Besides the realization of background turbulence as a colored-in-time stochastic source of excitation of the linearized dynamics, structural
perturbation of the linearized dynamical generator have also been sought which can be used to identify important state-feedback interactions that are lost through linearization. Such dynamical perturbations of the linearized equations can also be trained using partially observed statistical signatures of the turbulent flow using convex optimization; see~\cite{zarjovgeoCDC16,zargeojovARC20,zarmohdhigeojovTAC20} for additional details.

\subsection{Challenges and contribution}
\label{sec.probform}

\begin{figure}
\centering
\includegraphics[width=.15\textwidth]{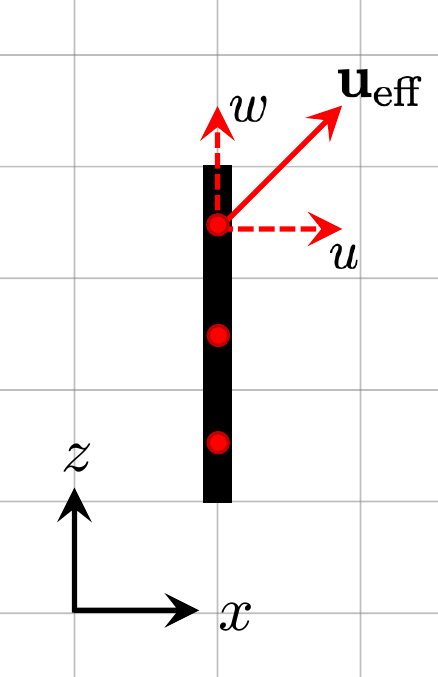}
\vspace{-.2cm}
    \caption{Geometric sketch of a 2D grid of collocation points around a turbine rotor. The sample grid demonstrates the division of the turbine rotor into 3 equally sized segments. The staggered points where the effective velocities $\bu_{\mathrm{eff}}$ and the intensities used in Equations~\eqref{eq.FP-expansions} are computed are marked by the red dots.}
    \label{fig.staggered-blade}
\end{figure}

One of the simplest representations of a wind turbine rotor is given by the actuator disk model (ADM), which assumes turbine rotors to be porous disks that experience uniform wind conditions over their entire surface and extract wind power by exerting a thrust force onto the incoming flow~\cite[Chapter 3]{burjenshabos11}. The energy extraction process within a control volume around the actuator disk yields the following expressions for the exerted thrust force and extracted power:
\begin{align*}
    F \;=\; \dfrac{1}{2}\,\rho\,A\,C_{T}\,\bu^2, 
    \quad\quad 
    P \;=\; \dfrac{1}{2}\,\rho\,A\,C_{P}\,\bu^3.
\end{align*}
Here, $F$ is the thrust force, $P$ is the power, $\rho$ is the air density, $A$ is the area of the rotor disk, $\bu$ is the effective surface-averaged velocity on the rotor disk, and $C_T$ and $C_P$ are, respectively, thrust and power coefficients that can be defined as functions of the axial induction factor $a$ that is used to quantify the induced flow at the actuator disk due to pressure variation over the turbine:
\begin{align*}
    C_T \;=\; 4a \left( 1 \,-\, a \right), 
    \quad\quad 
    C_P \;=\; 4a \left( 1 \,-\, a \right)^2.
\end{align*}
The maximum value of $C_p$, which characterizes the Betz limit for turbine efficiency, is obtained with $a = 1/3$ as $0.593$~\cite{burjenshabos11}. However, physical testing of turbines typically yields efficiency levels that are lower than this theoretically established bound. Following a similar approach as the ADM with rotation from Port{\'e}-Agel et al.~\cite{porluwu10} and Wu and Port{\'e}-Agel~\cite{wupor11}, the thrust force and power can be computed as the aggregate of contributions from constituting segments of a turbine rotor resulting from the discretization of the spatial domain, i.e., $F = \sum_i F_i$ and $P = \sum_i P_i$, where
\begin{align}
\label{eq.FP_segmented}
    F_i \;=\; \dfrac{1}{2}\,\rho\,A_i\,C_{T}\,\bu_{\mathrm{eff},i}^2, 
    \quad\quad 
    P_i \;=\; \dfrac{1}{2}\,\rho\,A_i\,C_{P}\,\bu_{\mathrm{eff},i}^3.
\end{align}
Here, $A_i$ represents the area of the rotor disk segment and $\bu_{\mathrm{eff},i}$ represents the effective velocity over the $i$th segment, which may be computed as the resultant velocity field evaluated on a staggered grid; see \textbf{Figure~\ref{fig.staggered-blade}} for an illustration.

\begin{figure}[b!]
\centering
\includegraphics[width=.48\textwidth]{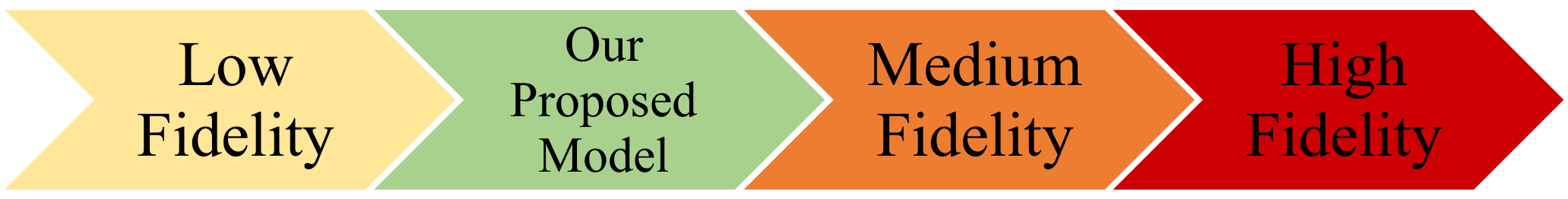}
    \caption{Our proposed modeling approach uses data to augment the predictive capability of low-fidelity engineering models using a stochastic dynamical representation of atmospheric turbulence whereby improved predictions of power and thrust force can be achieved.}
    \label{fig.ModelClas}
\end{figure}

\vsp
Low-fidelity analytical models such as the Frandsen model~\cite{frandsen06} or the Jensen-Park model~\cite{jen83,katic86} neglect the time-varying near-field turbulence behind the wind turbine and are often combined with linear wake superposition laws to provide an over-simplified prediction of wake velocities under steady atmospheric conditions. In the absence of a turbulence model that can capture the effect of the ABL and rotor-induced mixing, velocity deficits predicted by such models are typically over-predicted, and thus, lead to inaccurate predictions of the load and power (\textbf{Figure~\ref{fig.AnalyticalPredictions}}). In this paper, we take a step in compensating the shortcomings of low-fidelity models via reduced-order modeling of second-order statistics of the velocity field that are pertinent in the prediction of thrust force and power for various turbines using Equation~\eqref{eq.FP_segmented} or turbulence intensities in accordance with field measurements or LES results. To this end, we adopt the stochastic dynamical modeling framework of Zare et al.~\cite{zarchejovgeoTAC17,zarjovgeoJFM17,zargeojovARC20} to model the effect of background turbulence using linear dynamical models and improve the predictive capability of low-fidelity engineering models without adding to their dimensional complexity (\textbf{Figure~\ref{fig.ModelClas}}). The resulting data-enhanced models are of low-complexity and are thus convenient for conducting linear stochastic simulations. They are also well-suited for analysis and synthesis using tools from modern robust control as they provide an explicit linear state-space representation for the dynamics of velocity fluctuations in wind farms.

\subsection{Paper outline}
\label{sec.outline}

The remainder of the paper is organized as follows. In Section~\ref{sec.prob-formulation}, we formulate a problem that addresses the challenge of matching the thrust force and power generated across various turbines by accounting for the dynamics of velocity fluctuations around a static flow field predicted by a conventional engineering model. In Section~\ref{sec.SF-lnse}, we provide details on the stochastically forced linearized NS model, which we use to model the turbulent velocity field at the hub height of a wind farm. In Section~\ref{sec.modeling}, we summarize the stochastic modeling framework that we use to shape the forcing into the linearized NS equations and match SCADA data. In Section~\ref{sec.results}, we apply our approach to the problem of matching LES-informed quantities such as thrust force, generated power, and turbulence intensity across a multi-turbine wind farm and verify our results using linear stochastic simulations. We conclude with a summary of our contributions and potential future directions in Section~\ref{sec.remarks}.

\section{Problem formulation}
\label{sec.prob-formulation}

The wind velocity field $\bu$ in the farm can be decomposed into the sum of a time-averaged mean $\bar{\bu}$ and zero-mean fluctuations $\bv$ as 
\begin{equation}
\label{eq.MandFluct}
    \bu \;=\; \bar{\bu} \,+\, \bv, \quad\quad 
    \bar{\bu} \;=\; \bE \left[\bu \right],\quad \quad 
    \bE \left[\bv \right] \;=\; 0
\end{equation}
where overline and $\bE \left[ \cdot \right]$ both denote the time-average operator, e.g.,
\begin{align*}
    \bar{\bu}(\bx) 
    \;=\; 
    \bE \left[ \bu(\bx,t) \right] 
    \;=\; 
    \ds{\lim\limits_{T \to \infty} \dfrac{1}{T} \int^T_0 \, \bu(\bx,t \,+\, \tau)\, \mrd \tau}.
\end{align*}
Here, $\bx$ denotes the spatial coordinates and $t$ is time. The velocity fluctuation field $\bv$, which we will use to capture the effect of atmospheric turbulence on the wake model, is assumed to be a stochastic Gaussian process. When the velocity incident on the turbines is perpendicular to the rotor and there is no cross-wind, substitution of Equation~\eqref{eq.MandFluct} into Equation~\eqref{eq.FP_segmented} yields the following equations for the time-averaged thrust force and power associated with $i$th segment of the rotor:
\begin{subequations}
\label{eq.FP-expansions}
\begin{eqnarray}
    \label{eq.F-expansion}
    \bar{F}_i 
    &\,=\,&
    \dfrac{1}{2}\, \rho\,A_i\,C_{T} \left( \bar{\bu}_{\mathrm{eff},i}^2 \,+\, \overline{\bv_{\mathrm{eff},i}^2}
    \right)
    \\[.15cm]
    \label{eq.P-expansion}
    \bar{P}_i 
    &\,=\,&
    \dfrac{1}{2}\,\rho\,A_i\,C_{P} \left( \bar{\bu}_{\mathrm{eff},i}^2 \,+\, \overline{\bv_{\mathrm{eff},i}^2} \right)^{3/2},
\end{eqnarray}
\end{subequations}
where the effective velocities can represent the resultant of their components, e.g., $\bv_{\mathrm{eff}} = \sqrt{u^2 + w^2}$ when $\bv = [\,u\,~ w\,]^T$,
and the stochastic properties of the fluctuation field $\bv$, namely its zero mean (cf.~Equation~\eqref{eq.MandFluct}) and skewness (due to its Gaussian distribution), have been used to eliminate certain terms. Based on Equations~\eqref{eq.FP-expansions}, the scalar quantities that we obtain for the thrust force and power of each turbine are functions of not only the effective mean velocity $\bar{\bu}$, but also the  second-order statistics of the fluctuation field $\bv$ at the staggered points of the discretization grid. While analytical models provide a static prediction of the effective velocity in the wind farm (similar to $\bar{\bu}$), the fluctuation field $\bv$ provides an additional dynamic degree-of-freedom whose second-order statistics can be modeled to improve predictions of flow quantities across the farm; given a set of available time-averaged thrust force $\{\bar{F}_i\}$ and power $\{\bar{P}_i\}$ measurements for various turbines, the dynamics of $\bv$ can be sought to augment the predictions of static analytical models by providing the necessary second-order statistics $\overline{\bv_{\mathrm{eff}}^2}$ for matching the available data (cf.~Equations~\eqref{eq.FP-expansions}). On the other hand, the statistics of $\bv$ may be directly modeled to match turbulence intensities across the wind farm.

\vsp

A number of options exist for modeling velocity fluctuations $\bv$ including data-driven approaches. Herein, we follow the stochastic dynamical modeling approach of Zare et al.~\cite{zarchejovgeoTAC17,zarjovgeoJFM17,zargeojovARC20} and pursue stochastically forced linear time-invariant (LTI) approximations of complex wind farm flow dynamics. Specifically, we assume the following state-space representation 
\begin{align}
\label{eq.LTI-model}
    \ba{rcl}
    \bpsi_t(\bx,t)
	& =  &
	\bA\, \bpsi(\bx,t)
	\;+\;
	\bB\, \bd(\bx,t)
	\\[0.15cm]
	\bv(\bx,t)
	& = &
	\bC\, \bpsi(\bx,t)
	\ea
\end{align}
for the dynamics of velocity fluctuations $\bv$, where $\bpsi$ is the state vector, $\bd$ is a stationary zero-mean stochastic process, $\bA$ is the dynamic generator that represents the prior dynamical representation for the turbulent flow dynamics, $\bB$ is the input operator that is used to introduce the input $\bd$ into the dynamics, $\bC$ is an output operator that relates the state $\bpsi$ to the output velocity field $\bv$, and $(\cdot)_t$ is the partial derivative with respect to time. In this paper, we focus on physics-based dynamical approximations resulting from linearization of the NS around static base flow profiles that are generated by conventional engineering models. Nonetheless, alternative linear models, which may result from application specific assumptions/simplifications, or data-driven methods such as dynamic mode decomposition~\cite{sch10,jovschnicPOF14,annnicsei16,sch22} may also provide viable starting points for our modeling framework.
Together with the prior low-fidelity engineering model that predicts $\bar{\bu}$, the dynamical model considered for velocity fluctuations $\bv$ gives rise to a class of low-complexity models that are more accurate in predicting quantities that depend on turbulent flow statistics, but maintain a lower dynamic complexity relative to medium-fidelity models (\textbf{Figure~\ref{fig.ModelClas}}). 

\begin{figure}
    \centering
    \begin{tabular}{cccc}
         \subfigure[]{\label{fig.WindFarm}}
         && 
         \subfigure[]{\label{fig.VelocityField}}
         &
         \\[-.25cm]
         & 
         \begin{tabular}{c}
            \includegraphics[width=.28\textwidth]{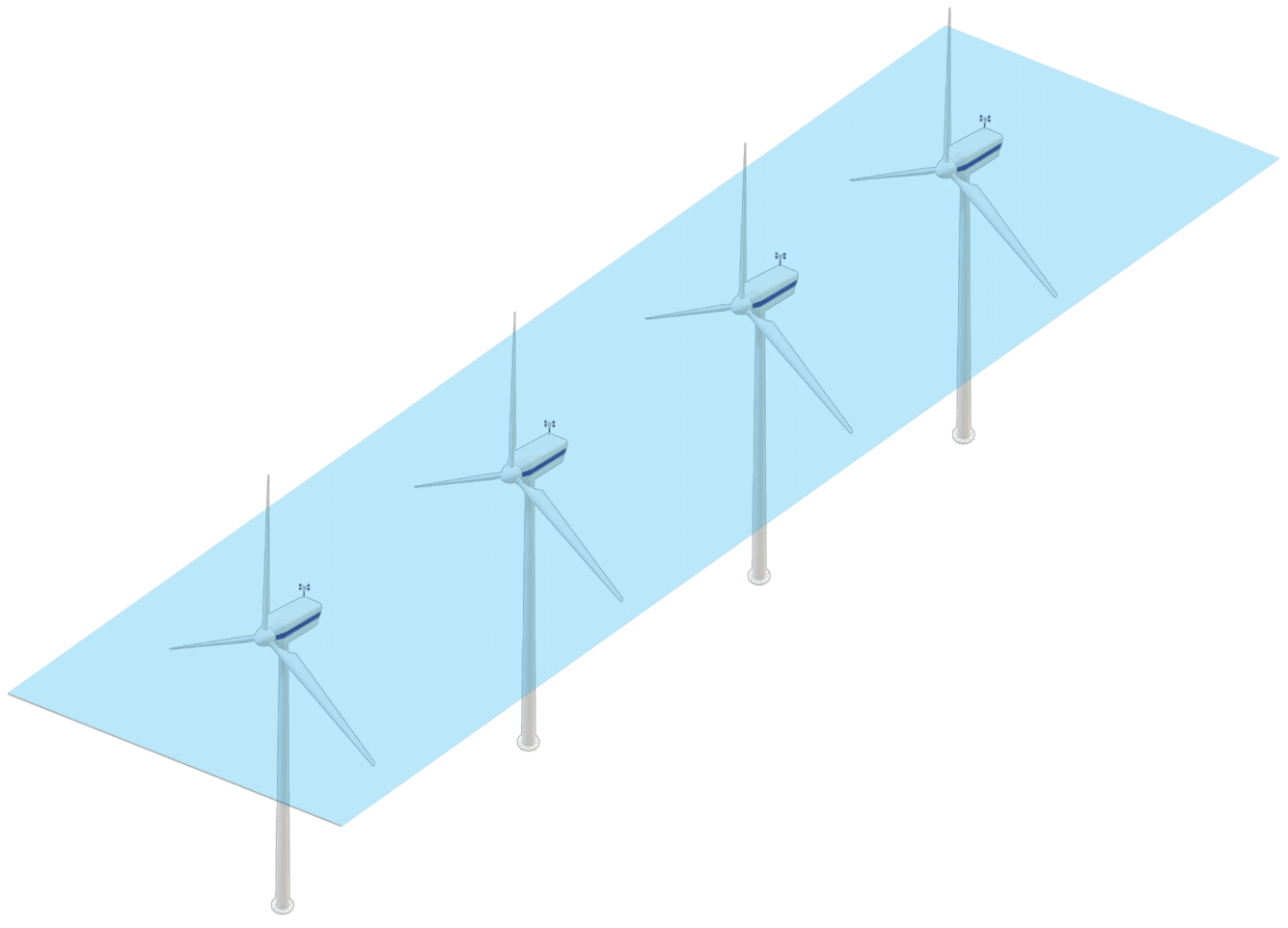}
         \end{tabular}
         &&
        \begin{tabular}{cc}
        \hspace{-0.2cm}
            \begin{tabular}{c}
            		\vspace{2cm}
            		\rotatebox{90}{\normalsize $z$}
            \end{tabular}
            &
            \hspace{-.4cm}
            \includegraphics[width=0.45\textwidth]{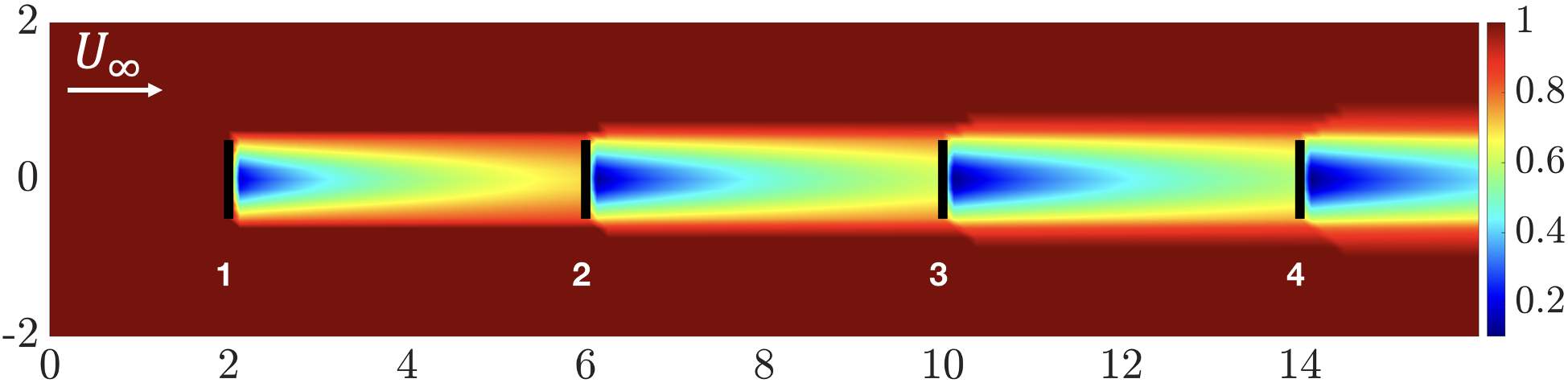}
            \\[-1.2cm]
            &
            \hspace{-.6cm}
            {\normalsize $x$}
        \end{tabular}
        \\[-.1cm]
        \subfigure[]{\label{fig.RawFcompare}}
         && 
        \subfigure[]{\label{fig.RawPcompare}}
         &
         \\[-.4cm]
         &
         \begin{tabular}{cc}
            \begin{tabular}{c}
		        \vspace{4cm}
		        {$\bar{F}$}
	        \end{tabular}
              & 
              \hspace{-.4cm}
             \includegraphics[width = 0.3\textwidth]{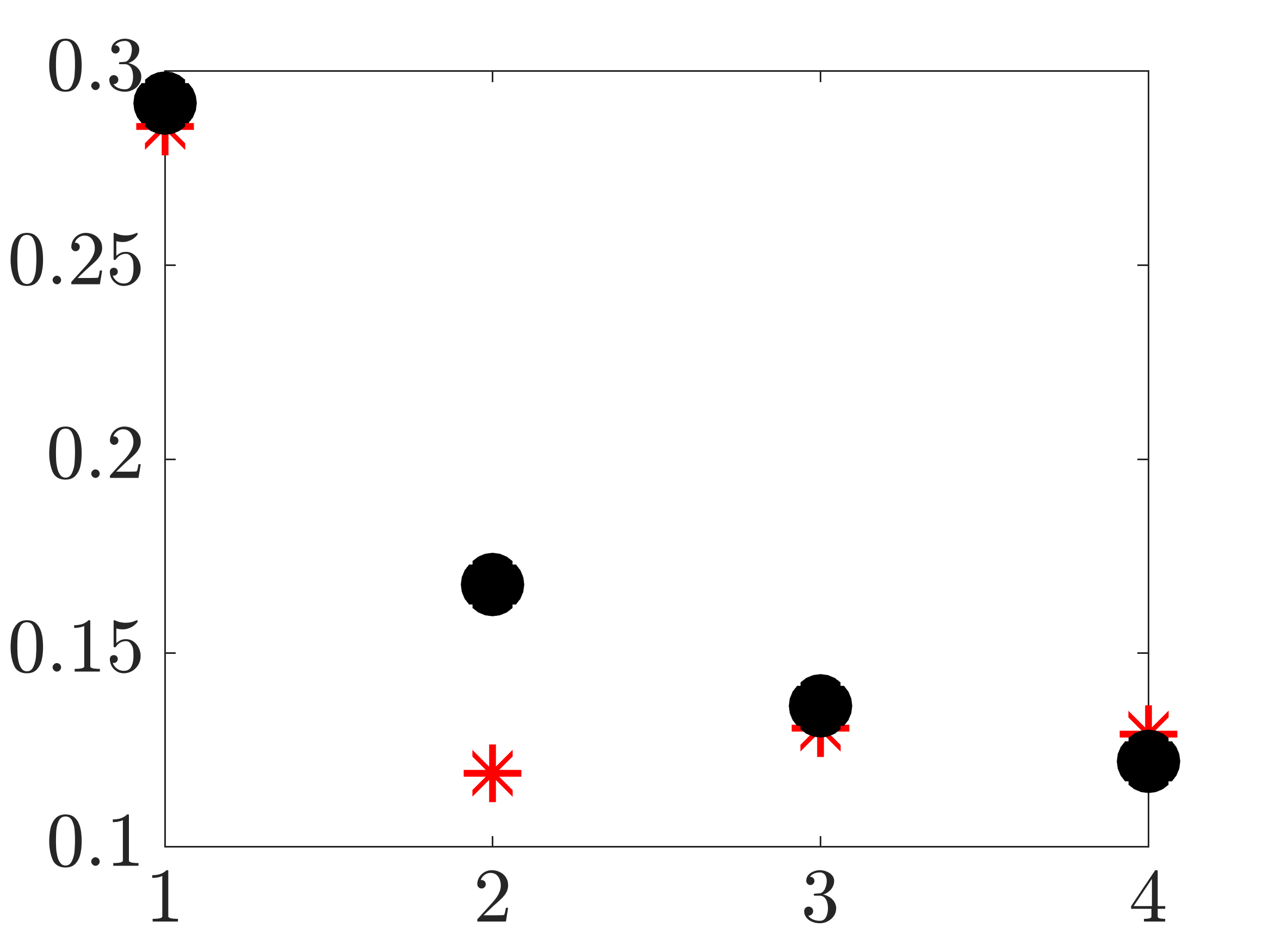} 
              \\[-2.1cm]
              & 
              turbine number
         \end{tabular}
         &&
         \hspace{-2.4cm}
         \begin{tabular}{cc}
            \begin{tabular}{c}
		        \vspace{4cm}
		        {$\bar{P}$}
	        \end{tabular}
              & 
              \hspace{-.4cm}
             \includegraphics[width = 0.3\textwidth]{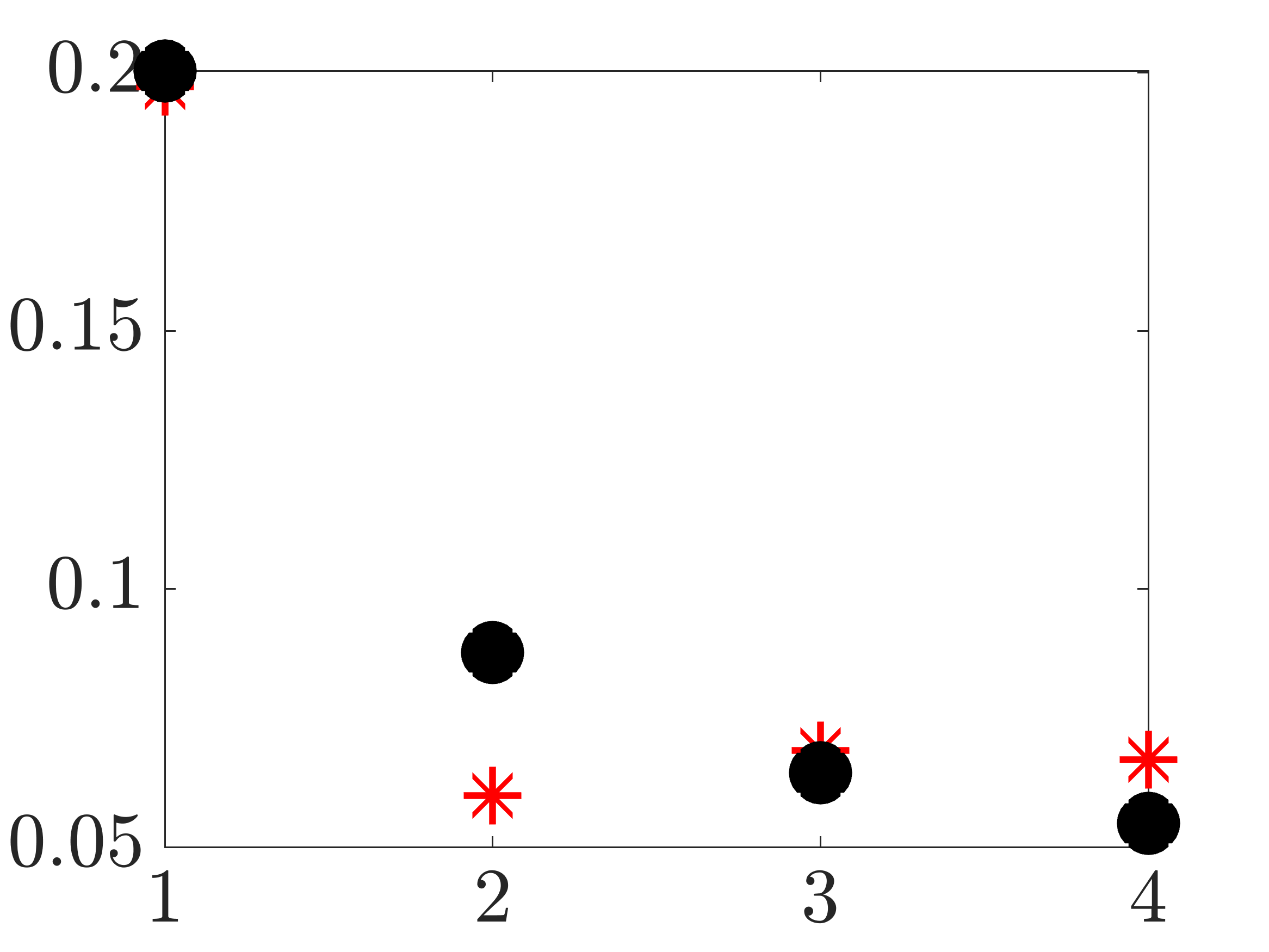}
              \\[-2.1cm]
              & 
              turbine number
         \end{tabular}
    \end{tabular}
    \caption{(a) A cascade of $4$ equally spaced turbines. (b) The streamwise and spanwise dependence of the velocity field $\bar{\bu}(x,z)$ generated by the analytical model~\eqref{eq.u-anal} from Bastankhah and Port{\'e}-Agel~\cite{baspor14} over the 2D computational domain at hub height. The thick black lines mark the location of the turbine rotors. (c) Predictions of thrust force $\bar{F}$ and (d) power generation $\bar{P}$ from LES (\textcolor{red}{{\large $\ast$}}) and the result of using Equation~\eqref{eq.FP_segmented} with $15$ segments across the spanwise extent of turbine rotors and the velocity field predicted by the analytical model~\eqref{eq.u-anal} ({\large $\bullet$}).}
    \label{fig.AnalyticalPredictions}
\end{figure}

\section{Stochastically forced linearized Navier Stokes equations}
\label{sec.SF-lnse}

In this section, we provide details on the stochastically forced linearized NS equations around a static 2D base flow profile $\bar{\bu}$ resulting from a low-fidelity engineering wake model. Our focus will be on 2D models of wind farm turbulence that are constrained to planes at the hub height of wind turbines. We note, however, that the proposed modeling framework is readily generalizable to 3D wind farm models that account for the remaining wall-normal dimension.

\vsp

The dynamics of small velocity and pressure fluctuations $(\bv,p)$ around the base flow profile $(\bar{\bu},\bar{P})$ are governed by the linearized NS and continuity equations
\begin{align}
\label{eq.lnse}
    \ba{rcl}
    \bv_t 
    &=&
    -\left(\nabla \cdot \bv \right)\bar{\bu} \,-\, \left(\nabla \cdot \bar{\bu} \right)\bv \,-\, \nabla p \,+\, \dfrac{1}{Re} \del \bv \,-\, K^{-1}\,\bv  \,+\, \bd
    \\[.15cm]
    0
    &=&
    \nabla \cdot \bv
    \ea
\end{align}
where the vector $\bv = [\,u~\,w\,]^T$, with $u$ and $w$ denoting components of fluctuating velocity field in the streamwise ($x$) and spanwise ($z$) directions, respective, $\nabla$ is the gradient operator, $\Delta=\nabla\cdot\nabla$ is the Laplacian operator, and the Reynolds number $Re=U_\infty d_0/\nu$ is defined in terms of the rotor diameter $d_0$, the free-stream velocity $U_\infty$, and the kinematic viscosity $\nu$. All variables in Equation~\eqref{eq.lnse} have been non-dimensionalized: length by $d_0$, velocity by $U_\infty$, time by $d_0/U_\infty$, and pressure by $\rho\, U_\infty^2$. In Equation~\eqref{eq.lnse}, $\bd$ represents an additive zero-mean stationary stochastic input that triggers a statistical response of the linearized dynamics. 

\vsp

\begin{figure}
    \centering
    \begin{tabular}{cccc}
        \subfigure[]{\label{fig.invKz}}
        &
        &
        \hspace{-.6cm}
        \subfigure[]{\label{fig.invK4x1}}
        &
        \\[-.2cm]
        &
         \hspace{-.9cm}
         \begin{tabular}{cc}
              \begin{tabular}{c}
                    \hspace{0.3cm}
            		\vspace{4cm}
            		\rotatebox{90}{\normalsize $K^{-1}$}
              \end{tabular}
              & 
              \hspace{-0.5cm}
              \includegraphics[width = 0.26\textwidth]{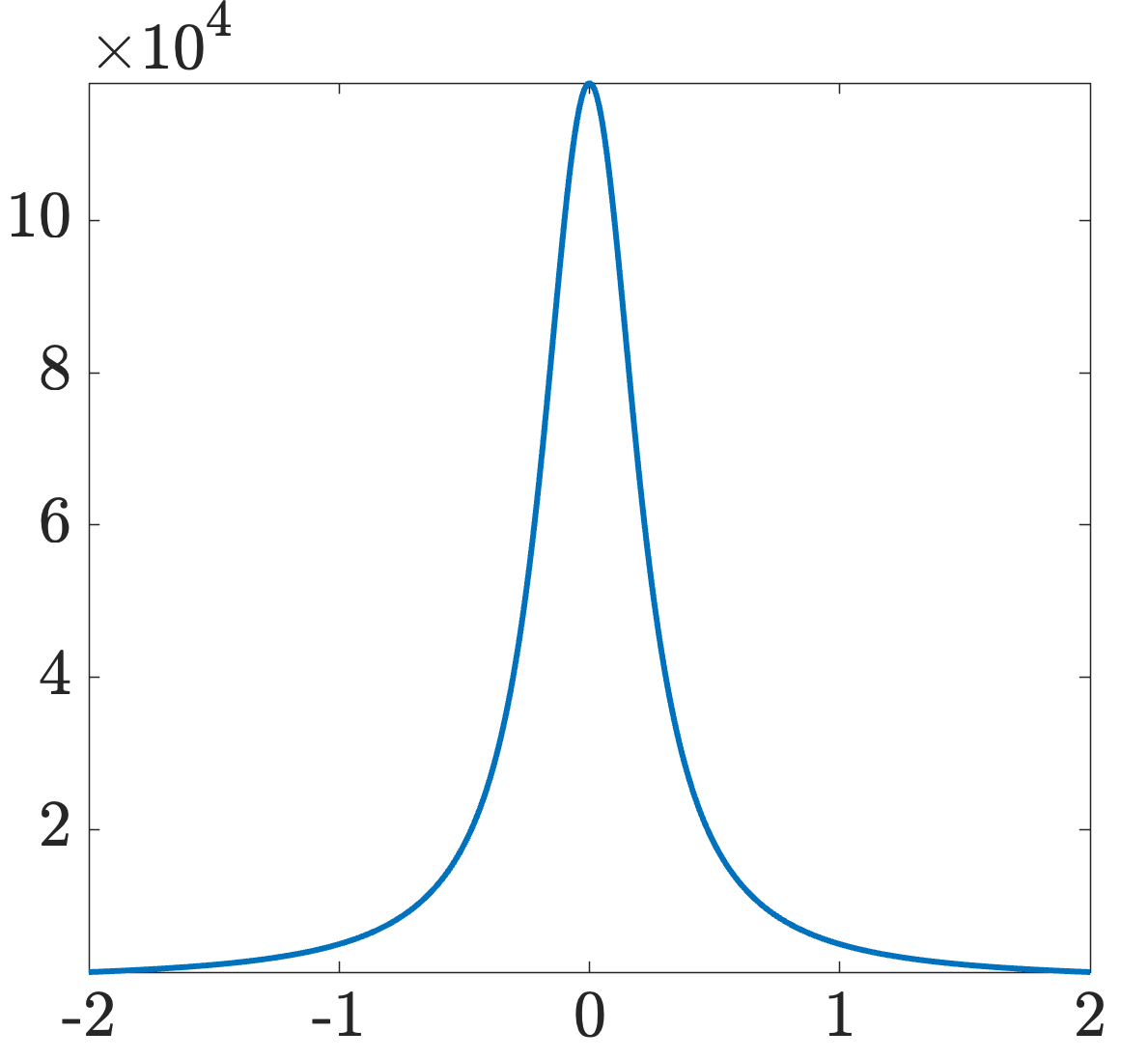}
              \\[-2.4cm]
              & 
              \hspace{-0.4cm}
              {$z$}
         \end{tabular}
         &&  
         \begin{tabular}{cc}
              \begin{tabular}{c}
            		\vspace{2.5cm}
            		\hspace{-0.8cm}
            		\rotatebox{90}{$z$}
              \end{tabular}
              & 
              \hspace{-0.8cm}
              \includegraphics[width = 0.62\textwidth]{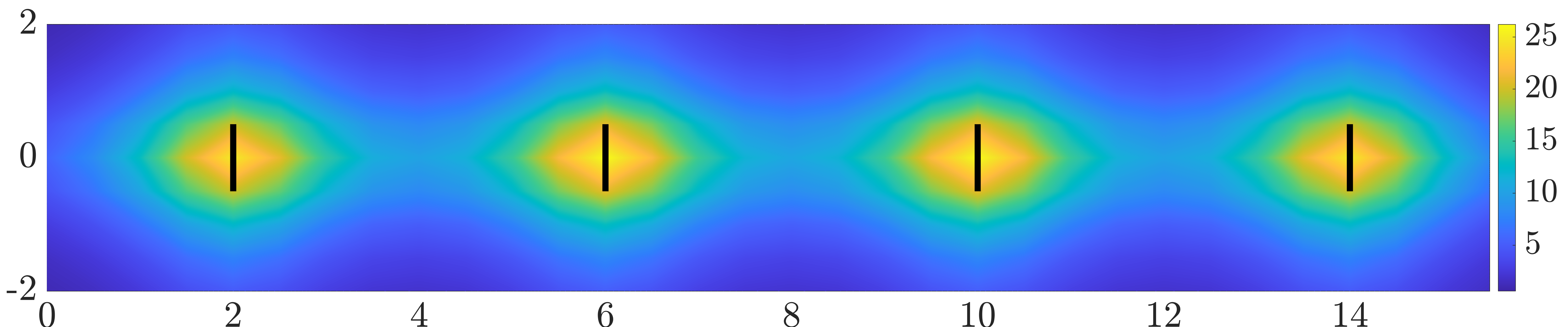}
              \\[-1.4cm]
              & 
              \hspace{-0.6cm}
              {$x$}
         \end{tabular}
    \end{tabular}
    \vspace{-0.4cm}
    \caption{(a) The spanwise dependence of the resistance function $K^{-1}(z)$ following Equation~\eqref{eq.invK} with $z_1 = -0.1$, $z_2 = 0.1$, $a = 5$, and $c = 400$. (b) The streamwise and spanwise dependence of the resistance function $K^{-1}(x,z)$ corresponding to Equation~\eqref{eq.invK}.}
    \label{fig.invK}
\end{figure}

In Equation~\eqref{eq.lnse}, the volume penalization term $K^{-1} \bv$ is used to capture the effect of turbine rotors and nacelles (and even turbine towers in 3D models) on the velocity field. This method avoids the implementation of boundary conditions in complex geometries by modeling the effect of solid obstructions of the flow as a spatially varying permeability function $K$ that influences the governing equations as an additive body force. Within the fluid, the penalization resulting from the permeability function $K$ should have no influence on the flow, i.e., $K \to \infty$, yielding back the original linearized NS dynamics for $\bv$. On the other hand, within solid structures, the function $K$ should force the velocity field to zero, i.e., $K \to 0$; see~\cite{khaangparcal00} for details. To capture the spatial region that is influenced by the presence of the turbines, we use a smooth 2D filter function of the form:
\vspace{.05cm}
\begin{align}
    \label{eq.invK}
    K^{-1}(x,z) \;=\; \dfrac{c}{\pi^2} \left[ \arctan(a(x\,-\, x_1)) \,-\, \arctan(a(x \,-\,x_2)) \right] \left[ \arctan(a(z\,-\,z_1)) \,-\, \arctan(a(z\,-\,z_2))\right],
\end{align}
where $x_{1,2}$ and $z_{1,2}$ determine the spatial extent of the rotors in the horizontal directions and parameters $a$ and $c$ determine the slope and magnitude of the function, respectively; see \textbf{Figure~\ref{fig.invK}} for samples of 1D and 2D resistance functions $K^{-1}$. Typically, the slope $a$ is set to a reasonably large value that clearly captures the spatial extent of the turbines but does not violate differentiability or cause large derivatives of $K^{-1}$ in the linearized operator. Ideally, the magnitude $c$ would be set to extremely large values to ensure a significant drop in the velocity field within the turbine structures. However, in practice, large values of $c$ can violate stability of the linearized NS operator. Our numerical experiments suggest that the largest value of $c$ before the linearized dynamics become marginally stable (eigenvalues of $A$ in Equation~\eqref{eq.LTI-model} fall on the imaginary axis) provides the best balance in capturing the effect of turbines on the turbulent velocity field while maintaining a well-behaved stable dynamic generator.

\vsp
A standard conversion for the elimination of pressure together with finite-dimensional approximation of the differential operators brings the linearized equations~\eqref{eq.lnse} into the form of the evolution model
\begin{align}
\label{eq.LTI-model-2}
    \dot{\bv}(t)
	\;= \;
	A\, \bv(t)
	\,+\,
	B\, \bd(t)
\end{align}
See Appendix~\ref{sec.appendixA} for the system matrices, details on the finite-dimensional approximation, and boundary conditions.
For the wind farm flow under steady atmospheric conditions, the global operator in Equation~\eqref{eq.LTI-model-2} has no exponentially growing eigenmodes (i.e., the dynamic matrix $A$ is stable). Thus, the steady-state covariance of the fluctuating velocity field
\begin{align*}
	X \,\DefinedAs\, \lim_{t\,\rightarrow \,\infty} \mathbf{E} \left( \bv(t)\, \bv^*(t) \right),
\end{align*}
subject to zero-mean white-in-time forcing $\bd$ with spatial covariance matrix $\Omega \succeq 0$, i.e., $\mathbf{E} \left( \bd(t)\right) = 0$ and 
$
    \mathbf{E} \left( \bd(t)\, \bd^*(\tau) \right) = \Omega\,\delta(t\,-\tau)
$
can be obtained from the solution to the Lyapunov equation,
\begin{align}
\label{eq.lyap}
	A\,X \,+\, X\,A^* 
	\;=\;
	-B\, \Omega\, B^*.
\end{align}
The Lyapunov equation~\eqref{eq.lyap} relates the statistics of white-in-time forcing, represented by $\Omega$, to the infinite-horizon state covariance $X$ via system matrices $A$ and $B$. The energy spectrum of the streamwise and spanwise velocity components can be extracted from the diagonal entries of the matrix $X$ and the total kinetic energy of the flow can be computed as $E = \trace(X)$. While white-in-time forcing is useful in studying the receptivity of the turbulent flow to exogenous disturbances~\cite{jovbamJFM05,ranzarhacjovPRF19b}, it is often found to be insufficient in reproducing its statistical signatures~\cite{zarjovgeoJFM17,zargeojovARC20}. To address this issue, we next consider the more general case of colored-in-time  stochastic forcing, and pose inverse problems that identify both the statistics of colored-in-time forcing and an input matrix $B$ to match available second-order statistics of wind farm turbulence using the LTI model~\eqref{eq.LTI-model-2}.

\section{Stochastic dynamical modeling of partially available second-order statistics}
\label{sec.modeling}

Modern-day wind farms use a host of sensing devices that are distributed across the farm to provide critical SCADA data for assessing the performance of the power plant and make changes to the operational settings of wind turbines in real time. The incoming stream of flow measurements from nacelle mounted anemometers, weather towers, pressure sensors, or even Doppler LiDAR systems can be processed to determine the power extracted by turbines, loads exerted on rotor structures, and the direction and speed of the incoming wind. Time averaged quantities can also be used to develop wake models that may in turn enable model-based flow estimation and wind farm control synthesis. Restricted by the modeling premise afforded by the segmented ADM model (Section~\ref{sec.prob-formulation}), herein, we utilize such data to realize stochastic forcing models for the linearized NS equations~\eqref{eq.LTI-model-2} that yield output velocity statistics that best reproduce the quantities of interest. We consider the availability of two types of data: (i) power and thrust force measurements at turbines; and (ii) velocity intensities at prespecified locations across the wind farm. While the second type of data (second-order statistics of the fluctuating velocity field $\overline{\bv^2}$) directly specifies entries in the covariance matrix $X$ of the linearized model~\eqref{eq.LTI-model-2}, the first type only provides such statistics through the ADM model; given time-averaged thrust force or power generation measurements across the farm, we use Equations~\eqref{eq.FP-expansions} together with a static approximation of an analytical wake model to obtain the resultant turbulence intensity at staggered points across the rotor structure and predict the competing quantity. Note that due to a lack of sufficient degrees of freedom in Equations~\eqref{eq.FP-expansions}, both thrust force and power measurements cannot be simultaneously matched. Details of how we obtain the turbulence intensity to match power or thrust force measurements or a balanced approximation of the two that addresses the issue of insufficient degrees of freedom are provided in Appendix~\ref{sec.appendixB}. Either of the three scenarios covered in the appendix yield an effective velocity intensity for each staggered point across the turbine rotors, but do not provide information regarding the contributions from different velocity components (e.g., $u^2$ or $w^2$), which may be provided via additional problem specific information such as the rotor yaw angle. Moreover, assuming knowledge of power and thrust force over individual segments of turbine rotors may not be a realistic expectation unless sensors are mounted on the surface of turbine blades. Consideration of scenarios where power or thrust force measurements are provided for entire turbines or turbine rotors are misaligned with the incoming wind direction is a topic of ongoing research. In Section~\ref{sec.MatchingFP}, we demonstrate how access to thrust force (power) measurements can results in predictions of power generation (thrust force) for wind turbines, and in Section~\ref{sec.TKE}, we demonstrate how partially observed second-order statistics of the velocity field can be used the complete the second-order statistical signature of the wind farm turbulence.

\vsp
Partially available second-order statistics of the velocity field $\overline{\bv^2}$ denote a subset of entries of the state covariance matrix $X$, which we wish to model. In the remainder of this section, we provide background material regarding the structural constraints on the state covariance matrix $X$, draw from the stochastic dynamical modeling framework of Zare et al.~\cite{zarchejovgeoTAC17,zarjovgeoJFM17,zargeojovARC20} to formulate covariance completion problems that identify the statistics of stochastic forcing $\bd$ into linear Gaussian model~\eqref{eq.LTI-model-2} to reproduce the available second-order statistics $\bv^2$, and provide details of a filter parameterization that enable the stochastic realization of the identified forcing.

\subsection{Second-order statistics of LTI systems}
\label{sec.stat-constraints}

For system~\eqref{eq.LTI-model-2} with Hurwitz $A$ and controllable pair $(A,B)$, a matrix $X$ qualifies as the steady-state covariance matrix of the state vector, i.e.,
\begin{align*}
	X \,\DefinedAs\, \lim_{t\,\rightarrow \,\infty} \mathbf{E} \left( \bpsi(t)\, \bpsi^*(t) \right),
\end{align*}
if and only if the Lyapunov-like equation
\begin{align}
\label{eq.lyap-like}
	A\,X \,+\, X A^* \;=\; -B\, H^* \,-\, H\,B^*
\end{align}
is solvable for the matrix $H$~\cite{geo02a,geo02b}. Here, $*$ denotes the complex conjugate transpose. The matrix $H$ quantifies the cross-correlation between the input and the state in model~\eqref{eq.LTI-model-2}~\cite[Appendix B]{zarjovgeoJFM17}:
\begin{align*}
    H \;\DefinedAs\; \ds{ \lim\limits_{t \to \infty} \, \bE \left[ \bpsi(t) \bd^*(t)\right] \,+\, \dfrac{1}{2} B\, \Omega}.
\end{align*}
When the stochastic input $\bd$ is zero-mean and white-in-time (state-independent) with covariance $\Omega$, $H = (1/2)B\, \Omega$ reduces Equation~\eqref{eq.lyap-like} to the standard algebraic Lyapunov equation~\eqref{eq.lyap}. In contrast to the Lyapunov Equation~\eqref{eq.lyap}, the right-hand side of Equation~\eqref{eq.lyap-like} is in general sign indefinite, i.e., will have both positive and negative eigenvalues unless the stochastic forcing $\bd$ is white-in-time.
The one-point velocity correlations along the diagonal of the state covariance matrix $X$ constitute turbulence intensities that are either matched in accordance with field measurements across the farm or model the deficits in matching thrust force or power generation in accordance with the segmented ADM model~\eqref{eq.FP-expansions}.

\subsection{Covariance completion}
\label{sec.cc-theory}

Given partially known diagonal entries of $X$ corresponding to deficits in matching thrust force, power generation, or turbulence intensities across the farm, we seek an input matrix $B$ and statistics of forcing $\bd$ that are consistent with the hypothesis that the required statistics in $\bv$ are generated by model~\eqref{eq.LTI-model-2} with known generator $A$. It is also important to restrict the complexity of the identified forcing model, which is quantified as the number of degrees of freedom that are directly influenced by the stochastic forcing, i.e., the number of input channels in matrix $B$ or $\rank(B)$. To these ends, we follow Zare et al.~\cite{zarchejovgeoTAC17,zarjovgeoJFM17,zargeojovARC20} in solving the structured covariance completion problem:
\begin{align}
	\hspace{-.03cm}
	\ba{cl}
	\minimize\limits_{X, \, Z}
	& 
	-\logdet\left(X\right) \,+\, \gamma\,\norm{Z}_*
	\\[.25cm]
	\subject 
	&
	~A \, X \,+\, X A^* \,+\, Z  \;=\; 0
	\\[.1cm]
	&
	\,\,X\circ E \,-\, G \;=\; 0
	 \ea
	\label{eq.CC}
\end{align}
which penalizes a composite objective subject to two linear constraints with the first corresponding to the Lyapunov-like equation~\eqref{eq.lyap-like} and the second denoting the set of known second-order statistics of the velocity field. Here, the matrices $A$, $C$, $E$, and $G$ are problem data, and the Hermitian matrices $X$, $Z$ are optimization variables. Entries of $G$ represent partially available second-order statistics of the velocity field $\bv$, the symbol $\circ$ denotes elementwise matrix multiplication, and $E$ is the structural identity matrix,
\begin{align}
	E_{ij} \;=\;
	\left\{
	\ba{ll}
	1,
	&~
	\text{if} ~ G_{ij} ~ \text{is available}
	\\[.1cm]
	0,
	&~
	\text{if}~ G_{ij}~ \text{is unavailable.}
	\ea
	\right.
	\non
\end{align}
The objective function provides a trade-off between the solution to a maximum-entropy problem and the complexity of the forcing model; the logarithmic barrier ensures the positive definiteness of the matrix $X$ and the nuclear norm regularizer, which is weighted by the parameter $\gamma>0$, is used as a proxy for the rank function (see, e.g., References~\cite{faz02,recfazpar10}). The rank of the matrix $Z$ bounds the number of independent input channels or columns in matrix $B$; for details see~\cite{zarchejovgeoTAC17}. We note that unless the forcing $\bd$ in Equation~\eqref{eq.LTI-model-2} is white-in-time, the matrix $Z$ may have both positive and negative eigenvalues. 
Convex optimization~\eqref{eq.CC} can be cast as a semidefinite program and solved efficiently using standard solvers~\cite{SDPT3,cvx,boyvan04} for small- and medium-size problems. In~\cite{zarjovgeoACC15,zarchejovgeoTAC17}, customized algorithms have been developed to deal with larger problems such as those arising in the modeling of multi-turbine wind farms.

\subsection{Stochastic realization}
\label{sec.filter-design}

\begin{figure}
	\begin{center}
        \includegraphics[width=0.55\textwidth]{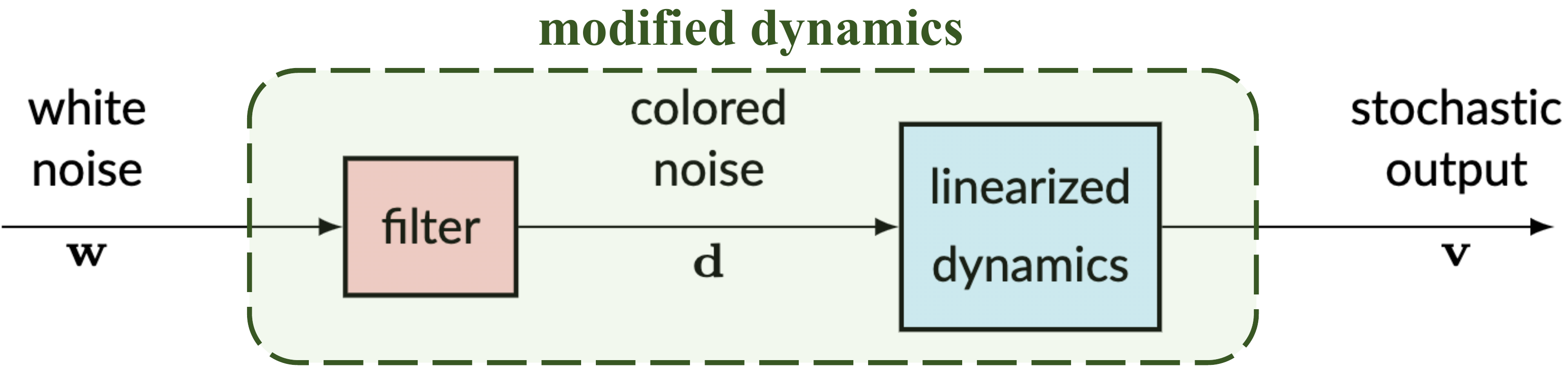}
	\end{center}
	\vspace{-.5cm}
	\caption{A cascade connection of an LTI system with a linear filter that is designed to account for the sampled steady-state covariance matrix $X$.}
	\label{fig.filter}
\end{figure}

Problem~\eqref{eq.CC} combines the nuclear norm with an entropy function in order to target low-complexity structures for stochastic forcing and facilitate the construction of a particular class of low-pass filters that generate suitable forcing into Equation~\eqref{eq.LTI-model-2}. The solution $Z$ to optimization problem~\eqref{eq.CC} can be decomposed into matrices $B$ and $H$ (cf.~Equation~\eqref{eq.lyap-like}) via spectral factorization. These factors, together with matrix $X$ that also results from solving problem~\eqref{eq.CC} and the state matrix $A$ enable the construction of generically minimal linear filters that have the same number of degrees of freedom as system~\eqref{eq.LTI-model-2} and are given by: 
\begin{subequations}
	\label{eq.filter}
	\begin{eqnarray}
		\dot{\xi}(t)
		&\!\! = \!\!&
		(A - B K)\, \xi(t)
		\;+\;
		B\, \bw(t)
		\\[0.05cm]
		\bd(t)
		&\!\! = \!\!&
		-K\, \xi(t)
		\;+\;
		\bw (t)
	\end{eqnarray}
where,
\be
	K 
	\;=\;
	\dfrac{1}{2}\, \Omega\,B^* X^{-1}
	\;-\;
	H^* X^{-1}.
\ee
\end{subequations}
Here, $\xi$ is the state of the filter and $\bw$ is a zero-mean white-in-time stochastic process with covariance $\Omega \succ 0$; see \textbf{Figure~\ref{fig.filter}}. The minimal realization of the linear filter and linearized dynamics results in a parsimonious (low rank) modification to the original linearized dynamics (\textbf{Figure~\ref{fig.filter}}),
\begin{align}
    \label{eq.modified-dyn}
    \dot{\bpsi}(t) \;=\; \left( A \,-\, B\, K\right) \bpsi(t) \;+\; B\, \bw (t).
\end{align}                                                          
The resulting stochastic wake model is linear and maintains a close relation with the physics retained by the linearized NS equations due to the low-rank nature of the modification term $BK$. Thus, it is not only convenient for the purpose of conducting linear stochastic simulations and real-time model-based feedback control with provable performance guarantees, but it holds the promise to ensure satisfactory performance even when the real physical system deviates from the model used for design.

\section{Numerical experiment}
\label{sec.results}

In this section, we utilize the stochastic dynamical modeling framework presented in Section~\ref{sec.modeling} to account for partially available second-order statistics of the turbulent velocity field $\bv^2$ at the hub height of a wind farm. We begin with a brief discussion into the details of LES which were used to generate data for the training and verification/validation of our stochastic dynamical models. We demonstrate the capability of our models in improving the predictions of analytical wake models in capturing the thrust force and power generation over the turbines in the $4\times 1$ cascade shown in \textbf{Figure~\ref{fig.AnalyticalPredictions}}. We then focus on a single turbine configuration to assess the value of velocity statistics at various distances downstream of the turbine in training our data-enhanced stochastic wake model. We build on the results obtained from this case study to model the turbulent flow impinging on a cascade of 4 turbines. Finally, we provide a dynamical realization for the identified stochastic forcing and conduct linear stochastic simulations to verify the ability of our models in accounting for statistical signatures of wind farm turbulence.

\subsection{Large-eddy simulations}
\label{sec.LES}

A cascade of 4 NREL-5MW reference turbines~\citep{NREL-5MW} (\textbf{Figure~\ref{fig.WindFarm}}) is simulated using the LES code UTD-WF~\cite{santoni2015,santoni2017,Ciri17,ciri2018}, which employs the rotating ADM to account for the effect of rotating turbine blades and the immersed boundary method of Orlandi and Leonardi~\cite{Orlandi06} to account for the towers and nacelles.
The computational box extends $32\, d_0$, $10.24\, d_0$ and $10\, d_0$ in the streamwise, spanwise, and vertical directions, respectively. 
The distance between the inlet and the most upstream turbine is equal to $9\, d_0$.
No-slip conditions are applied at the bottom boundary of the computational domain in addition to the surfaces of nacelles and towers, free-slip conditions are applied at the top boundary,
periodic boundary conditions are imposed at the two spanwise sides, and radiative boundary conditions~\cite{orlanski1976} are implemented at the outlet.
The grid is stretched in the vertical direction in order to have a finer resolution in the regions where the turbine rotors are present; grid resolution in the refined sections with the turbines is uniform in all three directions, $\Delta x=\Delta z=\Delta y= 0.025\, d_0$.
Although the resolution is not sufficient to resolve the boundary layer flow around the tower accurately (as in most LES), the impermeability provided by the immersed boundary method reproduces blockage effects and overall momentum loss across the turbine structures. 

\vsp
In order to mimic the atmospheric boundary layer at the inlet, turbulence obtained from a precursor simulation is superimposed to a mean velocity profile expressed by the following law:
\begin{align}
\label{eq:ABL}
    \dfrac{U}{U_{\mathrm{hub}}}
    \;=\;
    \left(\dfrac{y}{y_{\mathrm{hub}}}\right)^\alpha,
\end{align}
where $U$ is the streamwise velocity component at height $y$, $U_{\mathrm{hub}}=U_{\infty}$ is the mean streamwise component of the wind velocity at hub height $y_{\mathrm{hub}}$, and $\alpha$ is the shear exponent, which we set to $\alpha=0.05$. 
The upstream velocity $U_{\infty}$ is chosen to be about $0.8\, U_{\mathrm{rated}}$. 
This allows using a standard region II control law\cite{johnson2006,laks2009} for the rotor dynamics where each turbine is assumed to extract the maximal energy from the incoming flow.
The precursor simulation is run in a computational box with periodic boundary conditions in both streamwise and spanwise direction, no-slip conditions at the bottom, and free-slip conditions at the top. 
Roughness cubes are placed on the ground (bottom of the computational domain) to enhance the generation of turbulence~\cite{leonardi2010}.
The superposition of the mean flow in Equation~\eqref{eq:ABL} and the turbulence from the precursor simulation results in a hub-height turbulence intensity of $8\%$ impinging the first turbine in the cascade. 
The time-averaged and root mean square profiles of velocity fluctuations are computed using $750$ instantaneous snapshots of the 3D velocity field generated by LES. The numerical experiments herein will consider a Reynolds number based on $d_0$ and $U_\infty$ equal to $Re=8\times 10^7$ in accordance with the LES.

\subsection{Base flow}
\label{sec.baseflow}

Our stochastic models are based on the stochastically forced linearized NS equations around a static base flow profile $\bar{\bu}$ with an analytical expression provided by a low-fidelity engineering wake model. 
For simplicity, we assume all turbines to be facing the wind, i.e., $0^\circ$ yaw angle relative to the free-stream velocity, restrict the computational domain to the 2D space at hub height (\textbf{Figure~\ref{fig.WindFarm}}), and assume zero cross-wind, which means that the base flow will only contain one non-zero component in the streamwise direction. For the base flow, we use the wake model proposed by Bastankhah and Port{\'e}-Agel~\cite{baspor14},
\begin{align}
\label{eq.u-anal}
    \bar{\bu}(x,z) 
    \;=\;
    U_\infty \,-\, U_\infty\left(1 \,-\, \sqrt{1 \,-\, \dfrac{C_T}{8 \left( k^\star x/d_0 \,+\, 0.2 \sqrt{\beta} \right)^2}} \right) {\large e}^ {\left( \,-\, \dfrac{1}{2\left( k^\star x/d_0 \,+\, 0.2 \sqrt{\beta} \right)^2} \left( \dfrac{z}{d_0} \right)^2 \right)},
\end{align}
where $d_0 = 1$ is the non-dimensional diameter of turbines and $k^\star=0.03$ is the wake growth rate, which we have chosen in accordance with earlier studies (e.g.,~\cite{baspor14},~\cite{zhaletiun20}). The choice of $C_P=0.485$ and $C_T=0.787$ correspond to the maximum power generated by a $5$MW NREL turbine~\cite{jonbutmussco09} using an LES code that leverages blade momentum element theory~\cite{santcarareleo17,sangarcirzhaiunleo20}. When considering multi-turbine farms, we follow a linear superposition law to capture velocity deficits in the overlapping regions where wakes interact. \textbf{Figure~\ref{fig.VelocityField}} shows the static 2D velocity field corresponding to Equation~\eqref{eq.u-anal} for a cascade of $4$ turbines, where we have used $\Delta_x=\Delta_z=0.125$ to discretize the horizontal dimensions.

\subsection{Thrust force and power predictions}
\label{sec.MatchingFP}

As shown in \textbf{Figures~\ref{fig.AnalyticalPredictions}(c,d)}, the monotonically decreasing velocity field predicted by the analytical model $\bar{\bu}$~\eqref{eq.u-anal} fails to capture the increase in the thrust force and power after the second turbine in the cascade shown in \textbf{Figure~\ref{fig.WindFarm}}. The monotonic decrease in the flow energy can be attributed to the absence of a turbulence model that can promote turbulence in the near wake of turbines thereby energizing the velocity field and subsequently leading to higher thrust force and power generation in downstream turbines. This issue is particularly evident in the predictions of power generation for turbines located toward the end of the cascade  (\textbf{Figure~\ref{fig.RawPcompare}}), indicating a deficiency that can only get worse in larger wind farms with more turbines. 

\vsp

\begin{figure}
    \centering
    \begin{tabular}{cccccc}
         \subfigure[]{\label{fig.MatchF}}
         &  
         & 
         \subfigure[]{\label{fig.MatchP}}
         &  
         & 
         \subfigure[]{\label{fig.MatchFnP}}
         &  
         \\
         \hspace{.2cm}
         \begin{tabular}{c}
		    \vspace{3.5cm}
		    {$\bar{F}$}
	    \end{tabular}
         &
        \includegraphics[width=4.6cm]{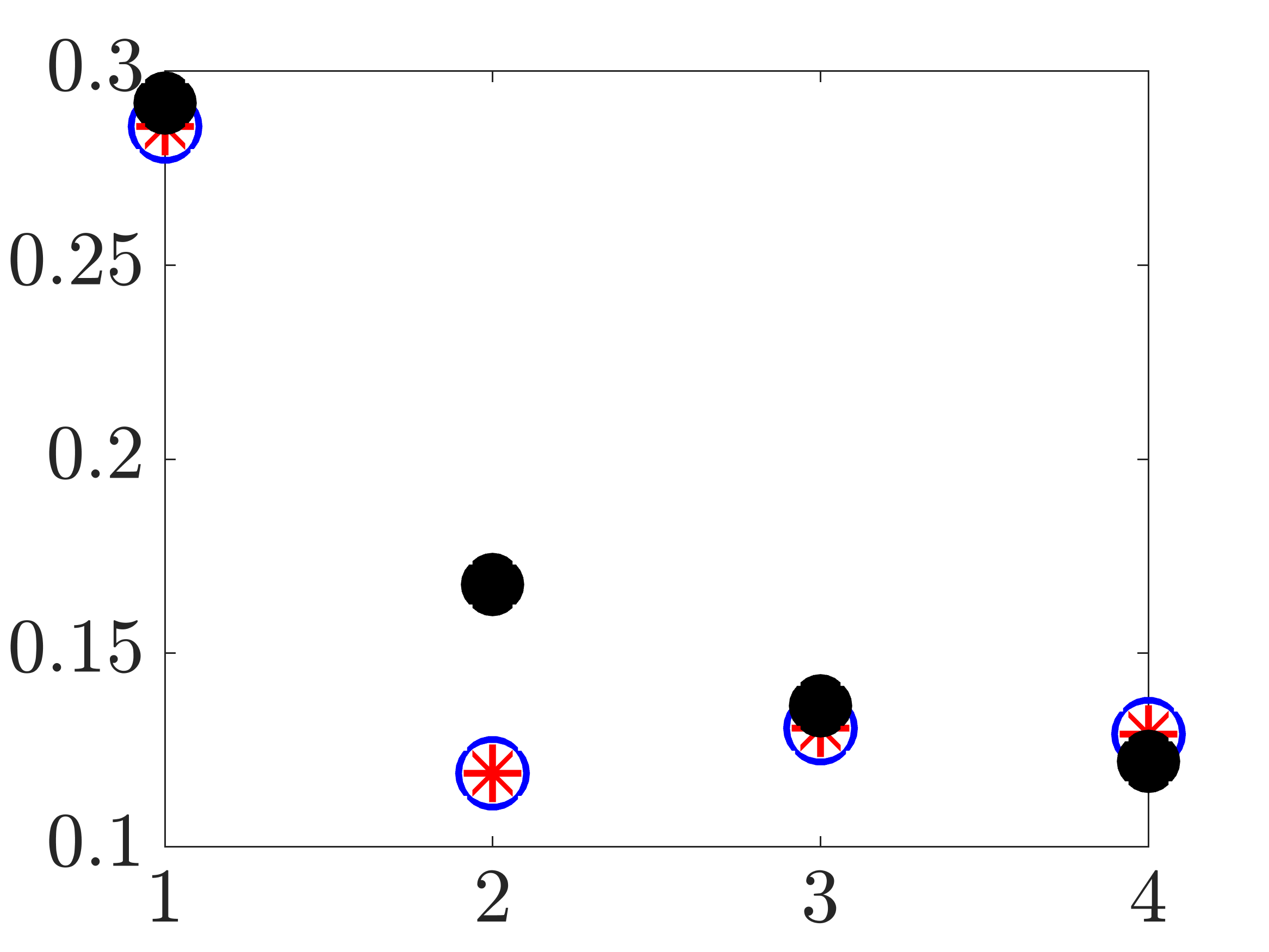}
         &
         &
        \includegraphics[width=4.6cm]{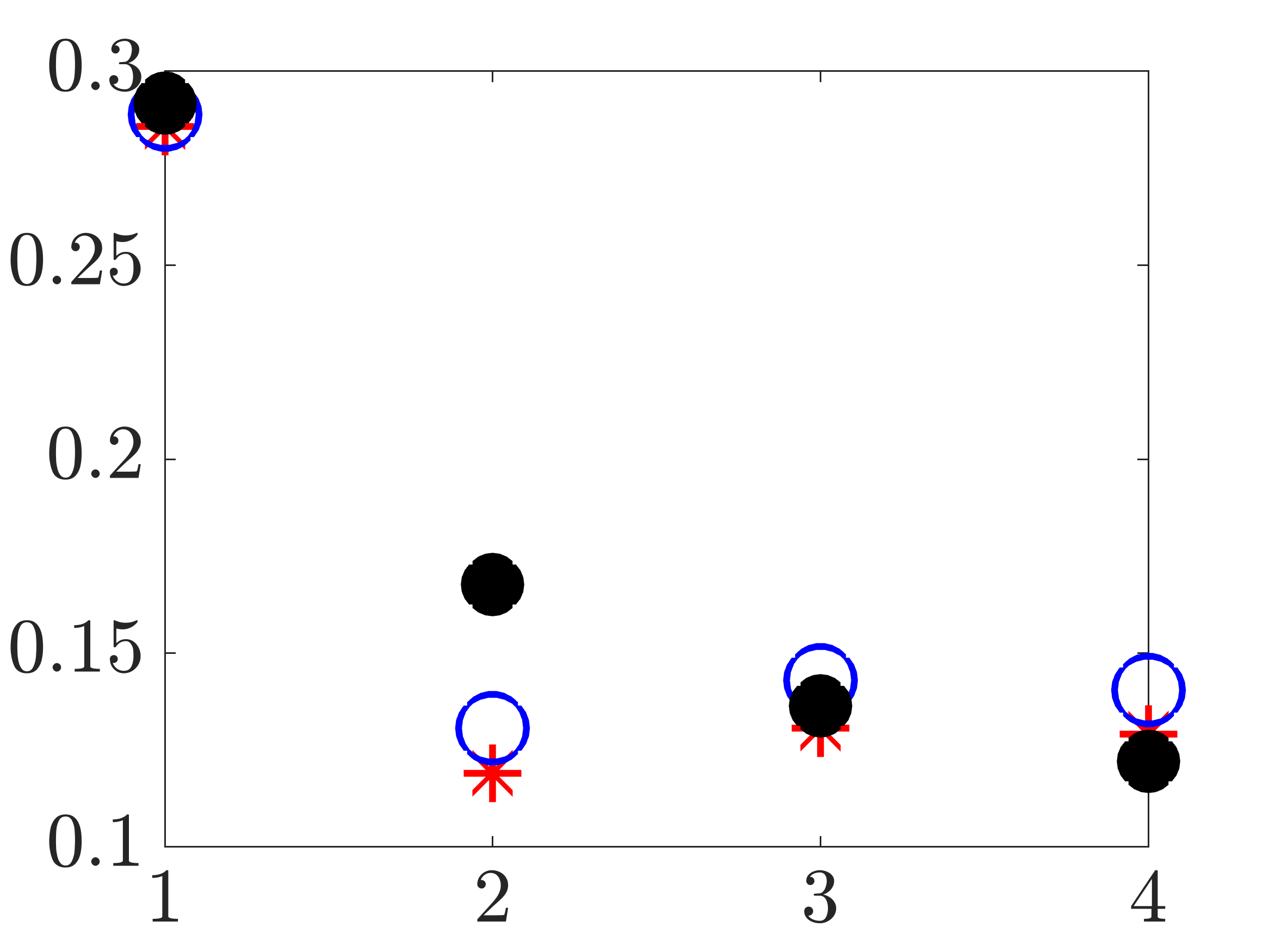}
         &
         &
        \includegraphics[width=4.6cm]{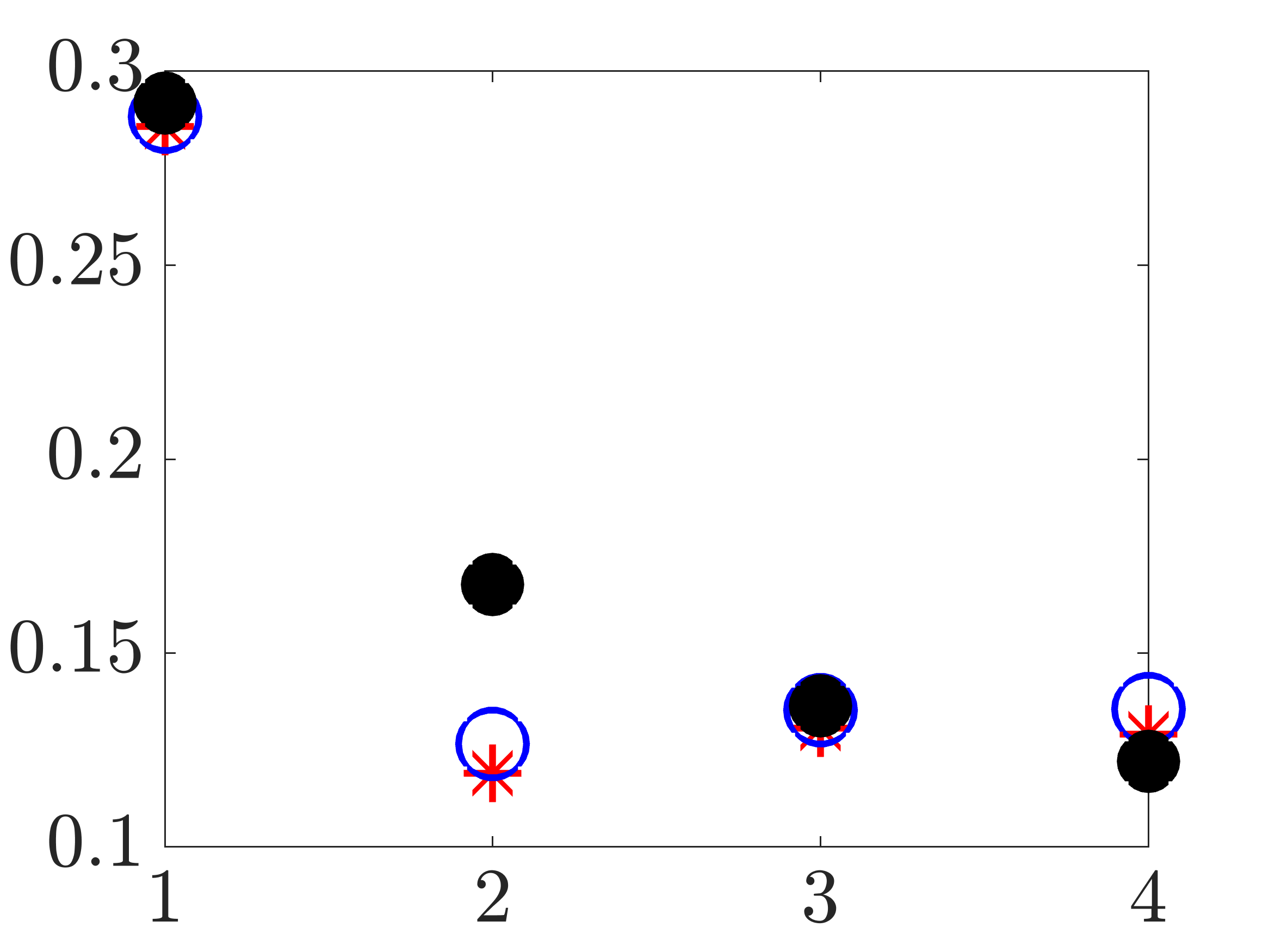}
        \\[-1.6cm]
        \hspace{.2cm}
        \begin{tabular}{c}
		    \vspace{3.5cm}
		    {$\bar{P}$}
	    \end{tabular}
         &
         \includegraphics[width=4.6cm]{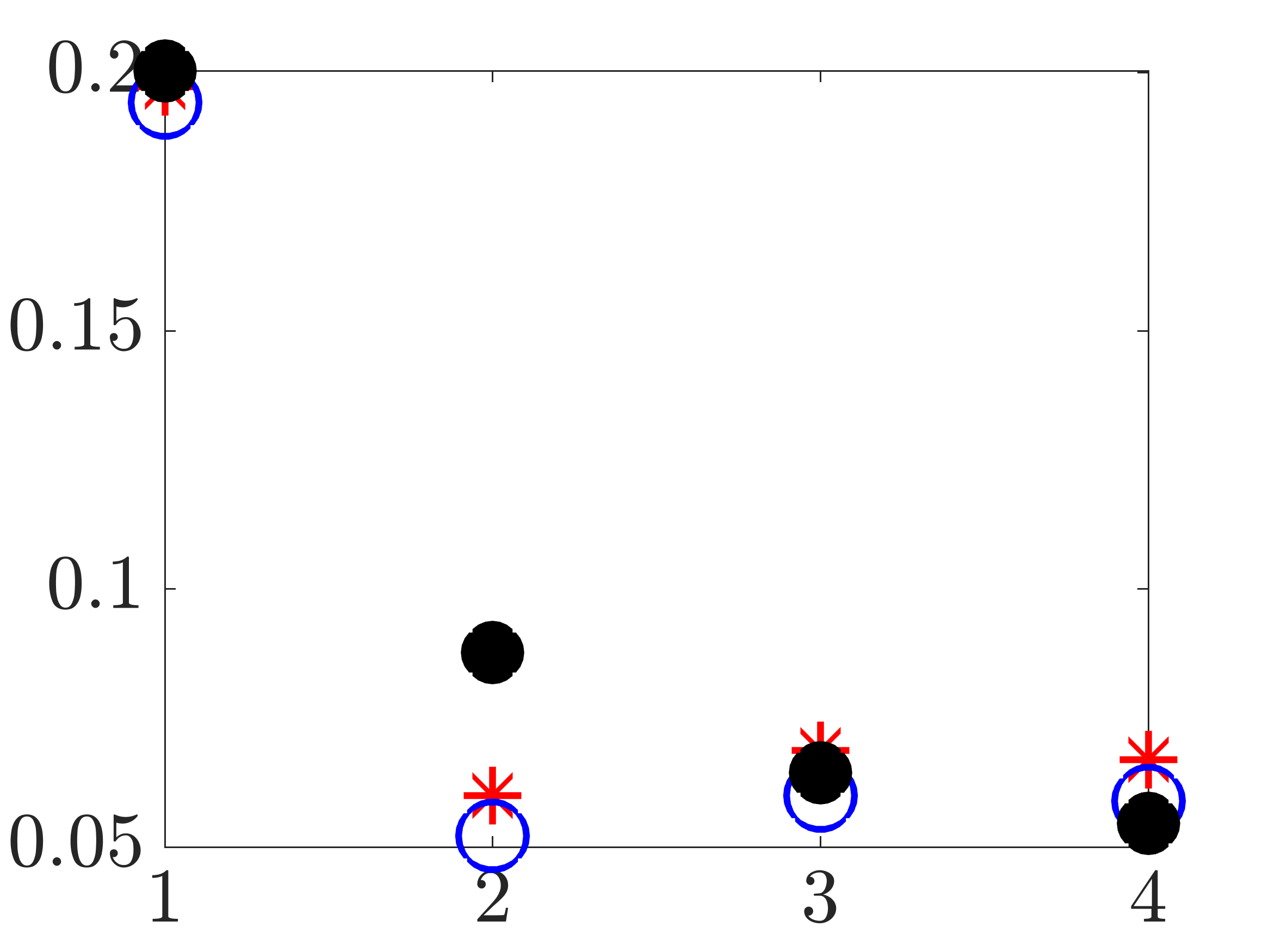}
         &
         &
        \includegraphics[width=4.6cm]{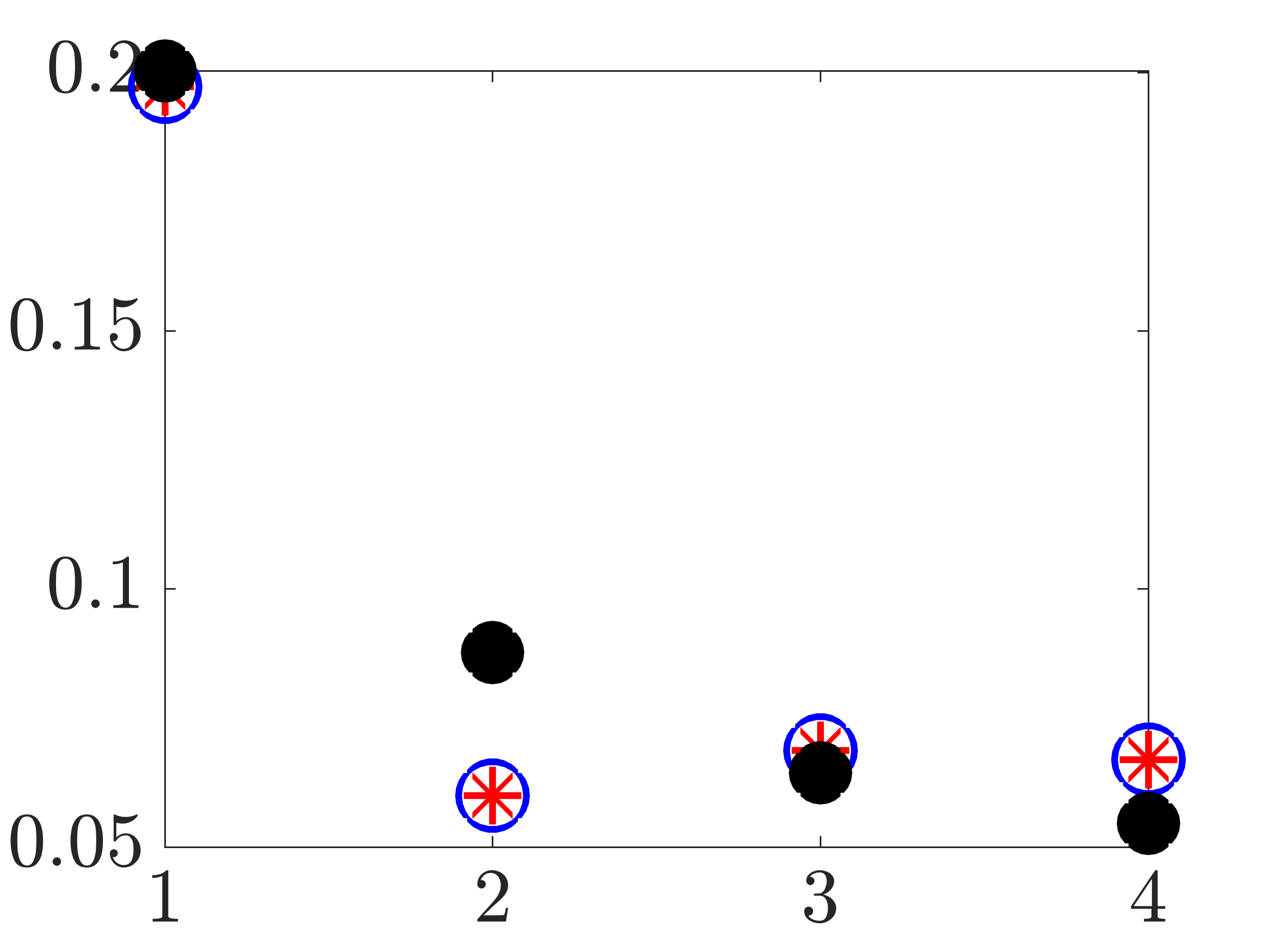}
         &
         &
        \includegraphics[width=4.6cm]{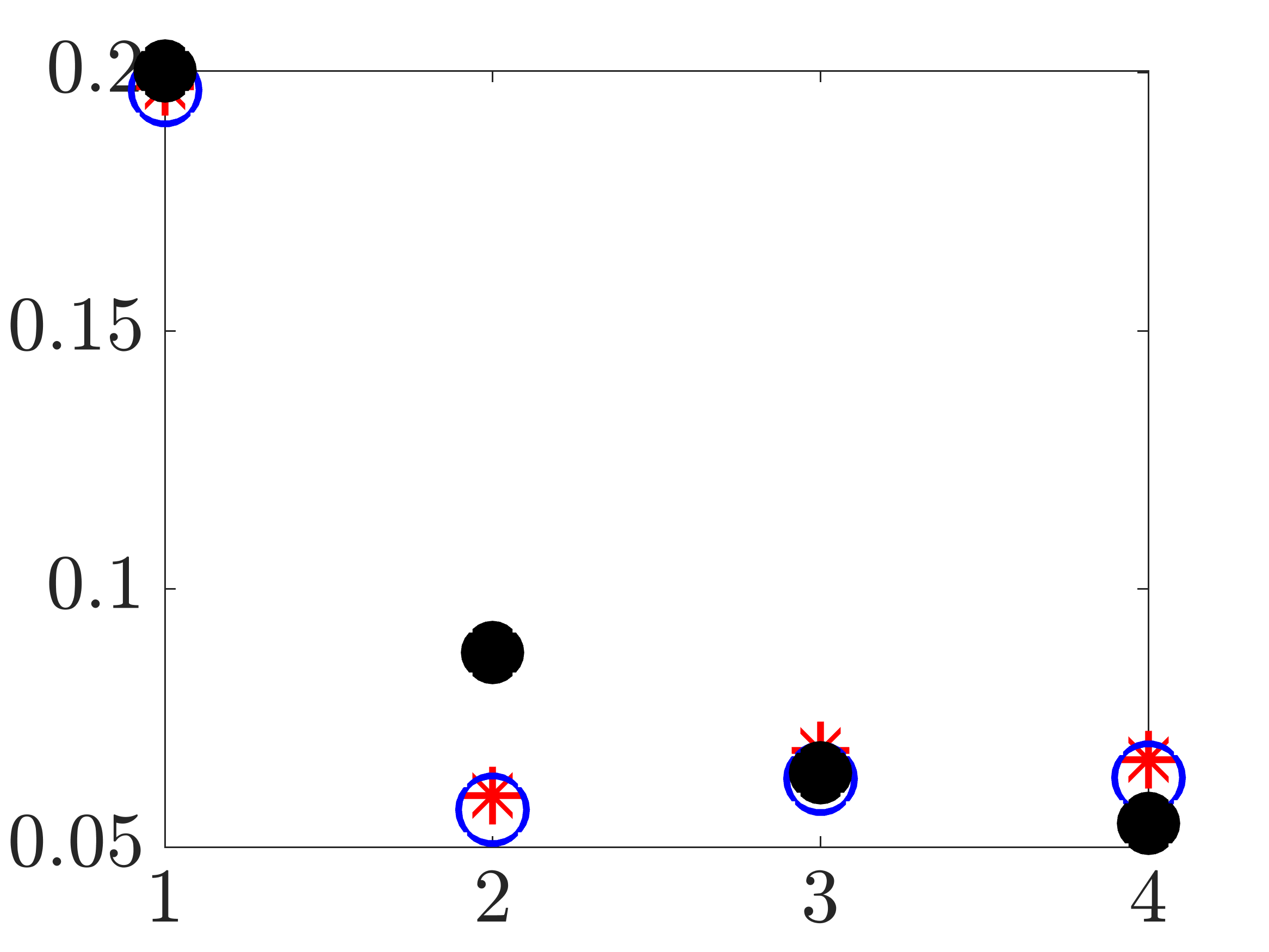}
        \\[-1.8cm]
        &
        \hspace{.4cm}
        {turbine number}
        &
        &
        \hspace{.4cm}        
        {turbine number}
        &
        &
        \hspace{.4cm}        
        {turbine number}
    \end{tabular}
    \vspace{-.2cm}
    \caption{(a) Results for matching thrust force $\bar{F}$ and  predicting power generation $\bar{P}$ over various turbines in the $4\times 1$ cascade; (b) Results for matching power generation $\bar{P}$ and predicting thrust force $\bar{F}$; (c) Results for matching the balanced approximation of both thrust force and power. LES data (\textcolor{red}{\large $\ast$}); predictions of of analytical model~\cite{baspor14} ({\large $\bullet$}); and predictions of our data-enhanced stochastic dynamical model (\textcolor{blue}{$\bigcirc$}).}
    \label{fig.MatchFP}
\end{figure}

To improve predictions of wake recovery, we model the statistics of velocity fluctuations $\bv$ around $\bar{\bu}$ using the linearized NS equations (Section~\ref{sec.SF-lnse}) subject to an optimally shaped source of additive stochastic excitation based on the the developments of Section~\ref{sec.modeling}. The velocity fluctuation field generated by the linearized NS equations augments the analytical wake model to improve predictions of thrust force and power generation based on the ADM model. \textbf{Figure~\ref{fig.MatchFP}} demonstrates that whether we match thrust force or power generation predictions of the competing quantity also improve. This improvement, 
which also captures the non-monotonicity of such quantities over the turbines in the cascade, 
is also observed when we match a balanced approximation of the two quantities of interest based on the solution to problem~\eqref{eq.least-squares} (\textbf{Figure~\ref{fig.MatchFnP}}). 
Importantly, the augmentation introduced to the predictions of the analytical model capture the non-monotonic trend of thrust force and power generation over the cascade of turbines. We anticipate this feature to be even more significant in larger arrays of wind turbines. Either way, we observe significantly improved predictions of the thrust force and power over the second, third, and forth turbines while predictions at the first turbine depreciate. This is perhaps due to the fact that we model turbulence intensities at the turbine locations per thrust force and power measurements, but do not explicitly account for the statistical signature of the incoming turbulence impinging on the array.

\subsection{Turbulence intensity predictions}
\label{sec.TKE}

In this section, we evaluate the predictive capability of our stochastic modeling framework in completing the statistical signature of the hub-height turbulent velocity field for a single turbine as well as a 4-turbine cascade of turbines. Specifically, we will assume knowledge of the streamwise and spanwise turbulence intensities at various diameters behind wind turbines and predict the remainder of the second-order statistical signature of the flow using the stochastically forced linearized NS model~\eqref{eq.LTI-model-2}. The available turbulence intensities may be provided by field measurement devices such as LiDAR systems that are deployed in wind farms to scan and monitor hub-height wind~\cite{iunwupor13} or may represent effective velocity intensities over rotor structures obtained from power and thrust force measurements (Appendix~\ref{sec.appendixB}). The optimization framework of Section~\ref{sec.cc-theory} identifies the appropriate colored-in-time forcing to the linearized NS equations to account for the available statistics and predict unavailable ones by virtue of the physics-based nature of model~\eqref{eq.LTI-model-2}.

\subsubsection{Predicting the wake of a single turbine using partially available flow intensities}
\label{sec.single-turbine}

\begin{figure}
    \centering
    \hspace{-0.5cm}
    \begin{tabular}{cccc}
         &  
         & 
         \hspace{-1.7cm}
         \subfigure[]{\label{fig.WF1x1_uuLES}}
         &
         \\
         &
         &
        \hspace{-1.5cm}
         \begin{tabular}{c}
		    \vspace{.5cm}
		    \rotatebox{90}{{$z$}}
	    \end{tabular}
         &
         \hspace{-1.5cm}
         \begin{tabular}{c}
            \includegraphics[width = 0.45\textwidth]{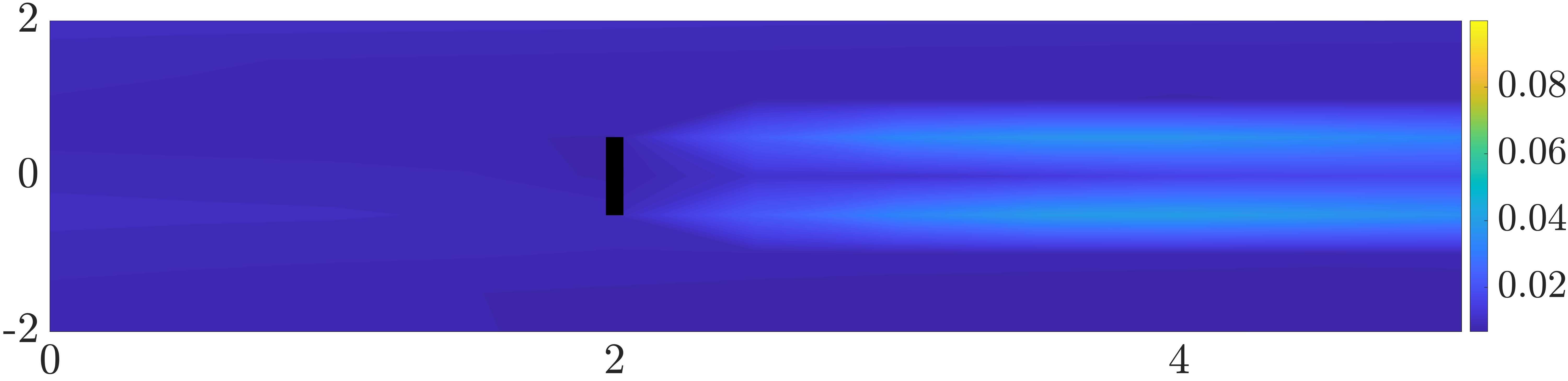}
         \end{tabular}
         \\
         \subfigure[]{}
         &  
         & 
         \hspace{-1.7cm}
         \subfigure[]{\label{fig.WF1x1_uuNM_1D}}
         &
         \\
         \begin{tabular}{c}
		    \vspace{.4cm}
		    \hspace{.3cm}
		    \rotatebox{90}{{$z$}}
	    \end{tabular}
         &
         \hspace{-0.6cm}
         \begin{tabular}{c}
            \includegraphics[width = 0.45\textwidth]{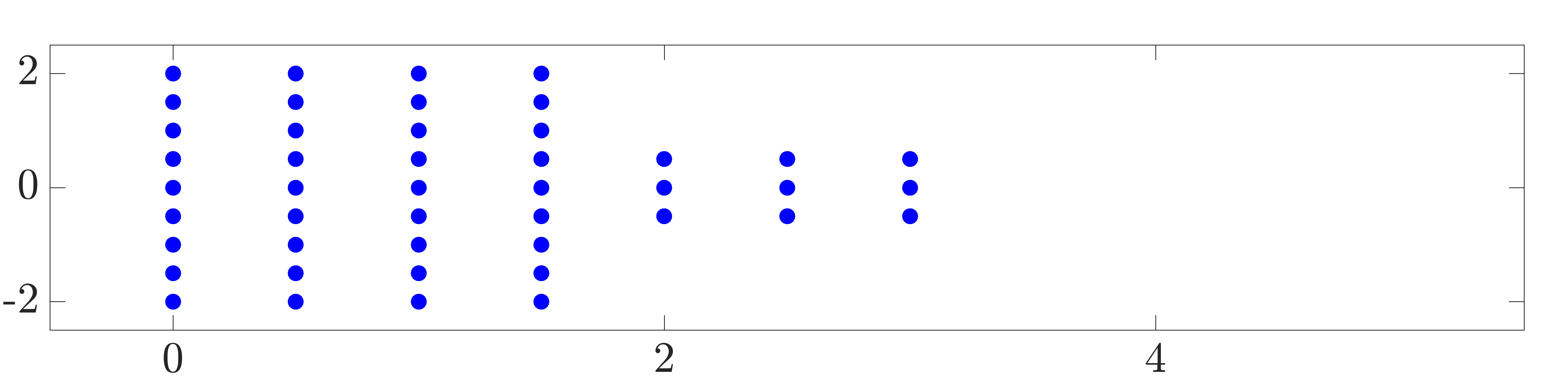}
         \end{tabular}
         &
         \begin{tabular}{c}
		    \vspace{.5cm}
		    \hspace{.5cm}
	    \end{tabular}
         &
         \hspace{-1.5cm}
         \begin{tabular}{c}
            \includegraphics[width = 0.45\textwidth]{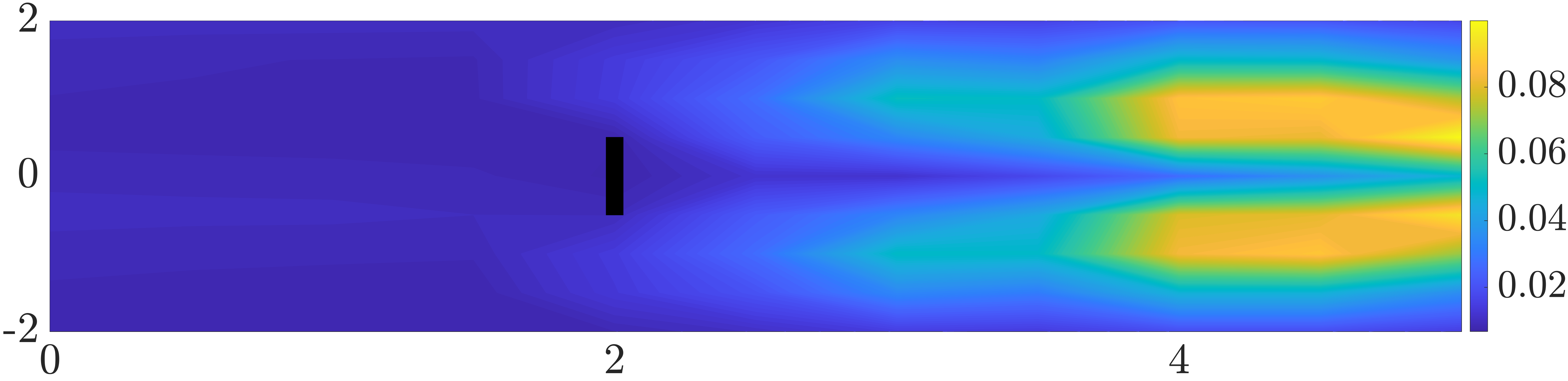}
         \end{tabular}
         \\
         \subfigure[]{}
         &  
         & 
         \hspace{-1.7cm}
         \subfigure[]{\label{fig.WF1x1_uuNM_2D}}
         &
         \\
          \begin{tabular}{c}
		    \vspace{.4cm}
		    \hspace{.3cm}
		    \rotatebox{90}{{$z$}}
	    \end{tabular}
         &
         \hspace{-0.6cm}
         \begin{tabular}{c}
            \includegraphics[width = 0.45\textwidth]{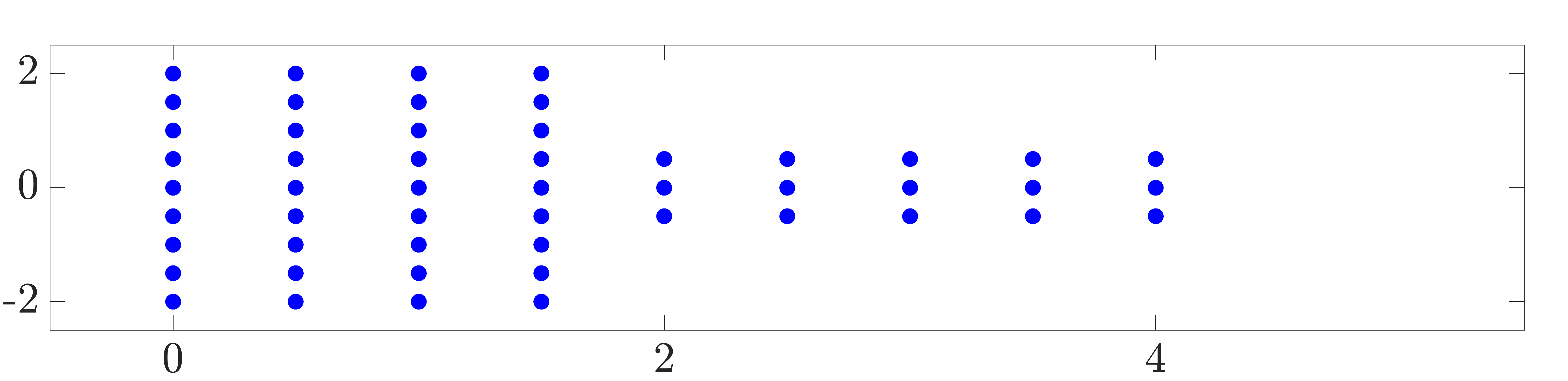}
         \end{tabular}
         &
         \begin{tabular}{c}
		    \vspace{.5cm}
		    \hspace{.5cm}
	    \end{tabular}
         &
         \hspace{-1.5cm}
         \begin{tabular}{c}
            \includegraphics[width = 0.45\textwidth]{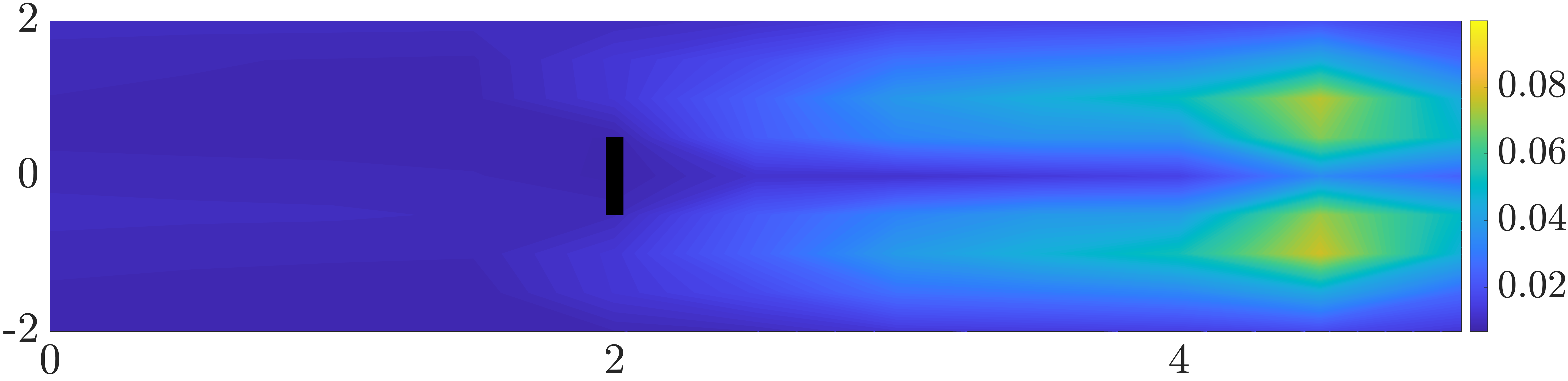}
         \end{tabular}
         \\
         \subfigure[]{}
         &  
         & 
         \hspace{-1.7cm}
         \subfigure[]{\label{fig.WF1x1_uuNM_3D}}
         &
         \\
          \begin{tabular}{c}
		    \vspace{.7cm}
		    \hspace{.3cm}
		    \rotatebox{90}{{$z$}}
	    \end{tabular}
         &
         \hspace{-0.6cm}
         \begin{tabular}{c}
            \includegraphics[width = 0.45\textwidth]{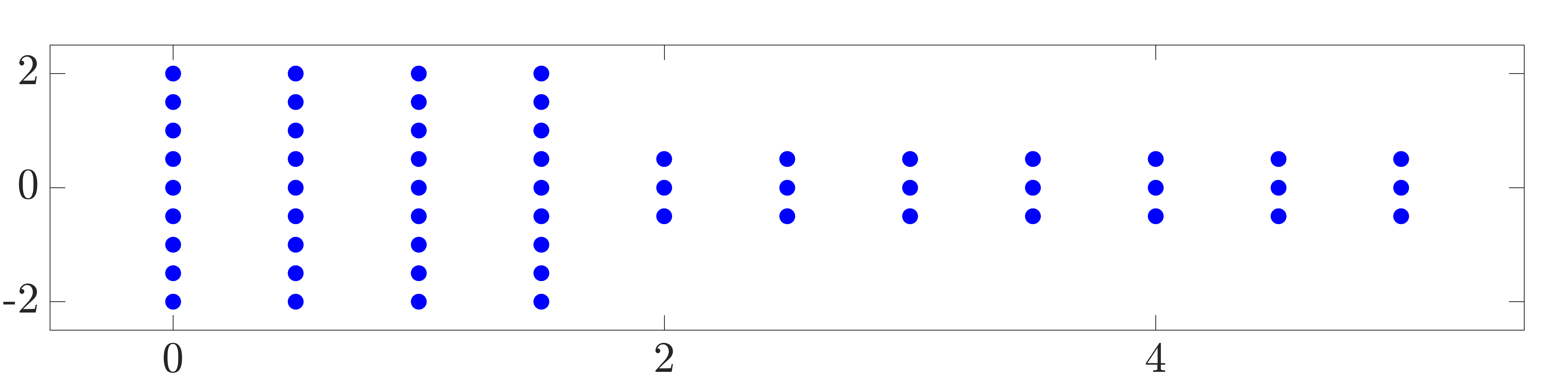}
            \\[-0.1cm]
            {$x$}
         \end{tabular}
         &
         \begin{tabular}{c}
		    \vspace{.5cm}
		    \hspace{.5cm}
	    \end{tabular}
         &
         \hspace{-1.5cm}
         \begin{tabular}{c}
            \includegraphics[width = 0.45\textwidth]{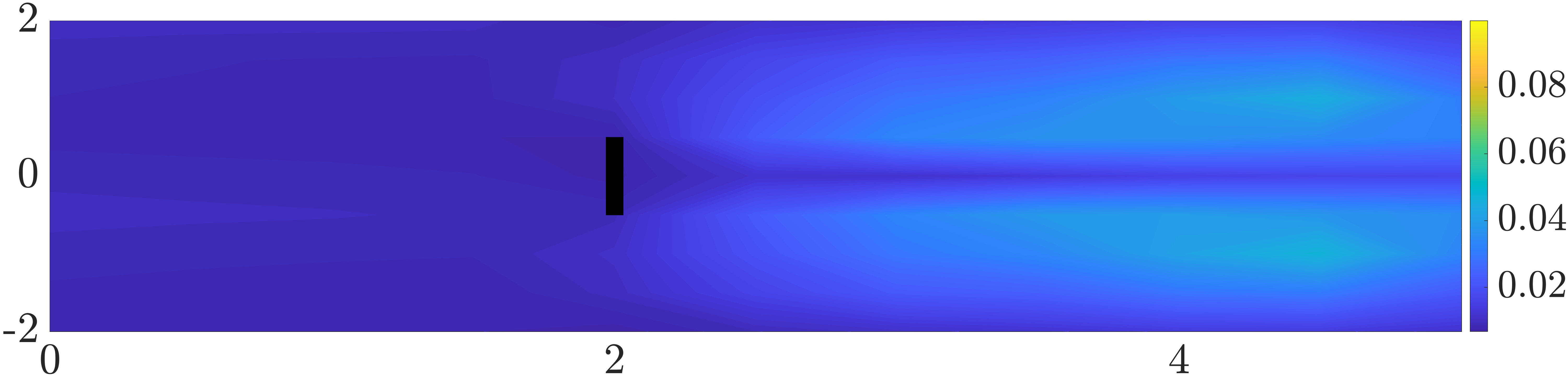}
            \\[-0.1cm]
            {$x$}
         \end{tabular}
    \end{tabular}
    \vspace{-.2cm}
    \caption{(a) Streamwise turbulence intensity ($uu$) obtained from LES and (c,e,g) the results of our stochastic dynamical model with data provided at $1 d_0$ (c), $2 d_0$ (e), and $3 d_0$ (g) locations downstream of the turbine as shown by the blue dots in the figures on the left.}
    \label{fig.WF1x1_matchuu}
\end{figure}

\begin{figure}
    \hspace{-0.5cm}
    \begin{tabular}{cccc}
         &  
         & 
         \hspace{-1.7cm}
         \subfigure[]{\label{fig.WF1x1_wwLES}}
         &
         \\
         &
         &
        \hspace{-1.5cm}
         \begin{tabular}{c}
		    \vspace{.5cm}
		    \rotatebox{90}{{$z$}}
	    \end{tabular}
         &
         \hspace{-1.5cm}
         \begin{tabular}{c}
            \includegraphics[width = 0.45\textwidth]{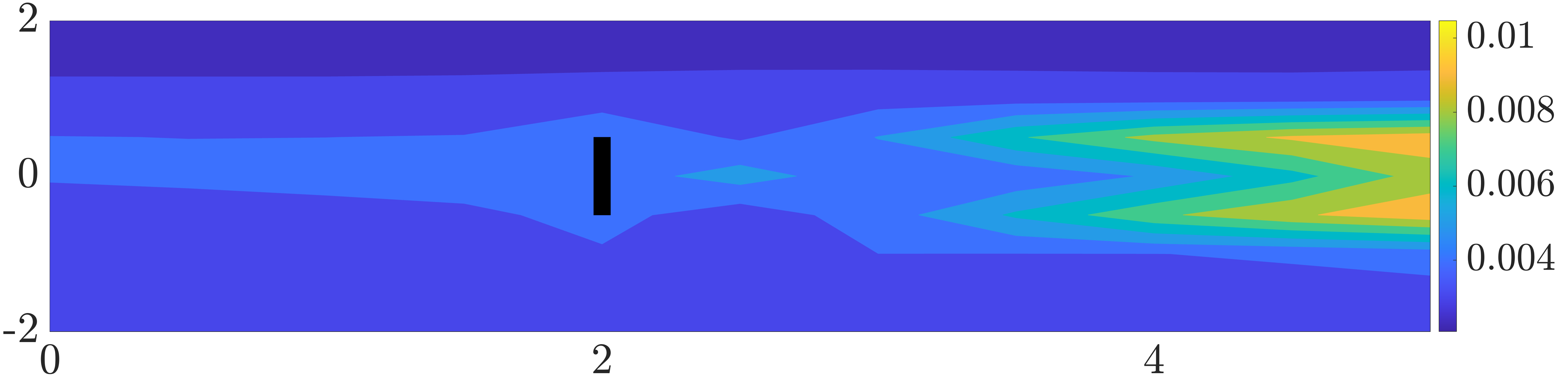}
         \end{tabular}
         \\
         \subfigure[]{}
         &  
         & 
         \hspace{-1.7cm}
         \subfigure[]{\label{fig.WF1x1_wwNM_1D}}
         &
         \\
         \begin{tabular}{c}
		    \vspace{.4cm}
		    \hspace{.3cm}
		    \rotatebox{90}{{$z$}}
	    \end{tabular}
         &
         \hspace{-0.6cm}
         \begin{tabular}{c}
            \includegraphics[width = 0.45\textwidth]{Gmat1D.png}
         \end{tabular}
         &
         \begin{tabular}{c}
		    \vspace{.5cm}
		    \hspace{.5cm}
	    \end{tabular}
         &
         \hspace{-1.5cm}
         \begin{tabular}{c}
            \includegraphics[width = 0.45\textwidth]{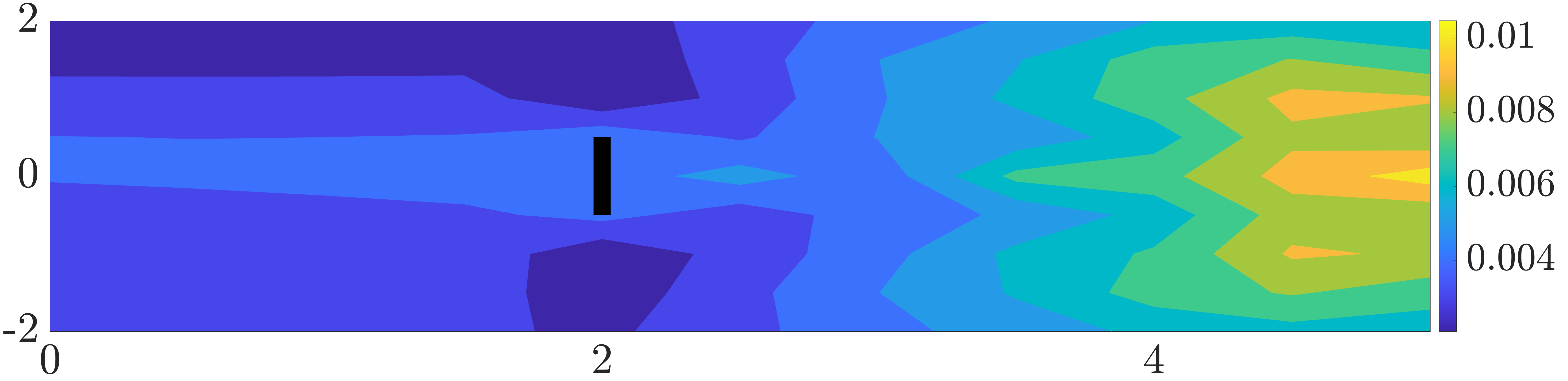}
         \end{tabular}
         \\
         \subfigure[]{}
         &  
         & 
         \hspace{-1.7cm}
         \subfigure[]{\label{fig.WF1x1_wwNM_2D}}
         &
         \\
          \begin{tabular}{c}
		    \vspace{.4cm}
		    \hspace{.3cm}
		    \rotatebox{90}{{$z$}}
	    \end{tabular}
         &
         \hspace{-0.6cm}
         \begin{tabular}{c}
            \includegraphics[width = 0.45\textwidth]{Gmat2D.png}
         \end{tabular}
         &
         \begin{tabular}{c}
		    \vspace{.5cm}
		    \hspace{.5cm}
	    \end{tabular}
         &
         \hspace{-1.5cm}
         \begin{tabular}{c}
            \includegraphics[width = 0.45\textwidth]{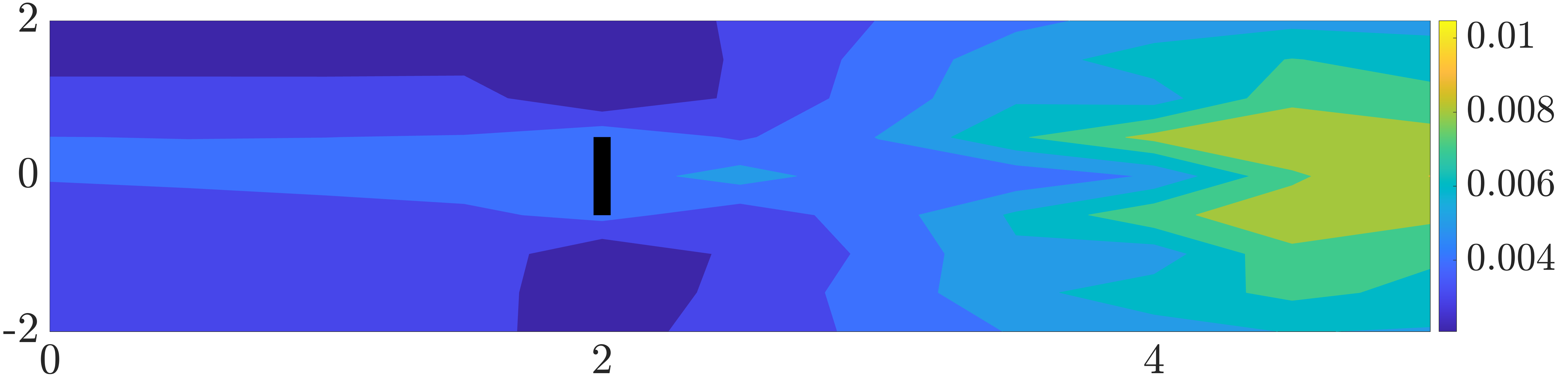}
         \end{tabular}
         \\
         \subfigure[]{}
         &  
         & 
         \hspace{-1.7cm}
         \subfigure[]{\label{fig.WF1x1_wwNM_3D}}
         &
         \\
          \begin{tabular}{c}
		    \vspace{.7cm}
		    \hspace{.3cm}
		    \rotatebox{90}{{$z$}}
	    \end{tabular}
         &
         \hspace{-0.6cm}
         \begin{tabular}{c}
            \includegraphics[width = 0.45\textwidth]{Gmat3D.png}
            \\[-0.1cm]
            {$x$}
         \end{tabular}
         &
         \begin{tabular}{c}
		    \vspace{.5cm}
		    \hspace{.5cm}
	    \end{tabular}
         &
         \hspace{-1.5cm}
         \begin{tabular}{c}
            \includegraphics[width = 0.45\textwidth]{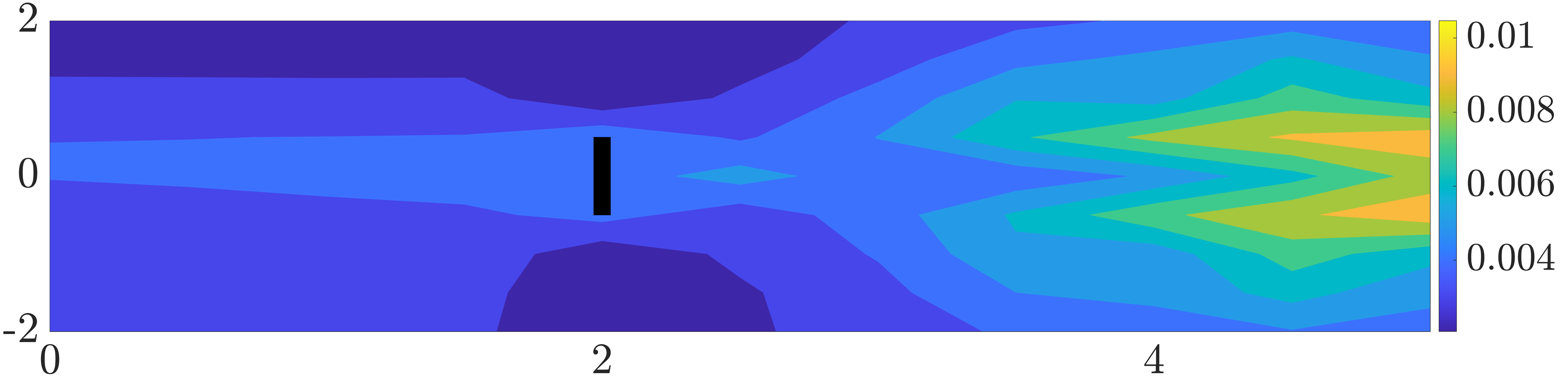}
            \\[-0.1cm]
            {$x$}
         \end{tabular}
    \end{tabular}
    \vspace{-.2cm}
    \caption{(a) Spanwise turbulence intensity ($ww$) obtained from LES and (c,e,g) the results of our stochastic dynamical model with data provided at $1 d_0$ (c), $2 d_0$ (e), and $3 d_0$ (g) locations downstream of the turbine as shown by the blue dots in the figures on the left.}
    \label{fig.WF1x1_matchww}
\end{figure}

We first focus on the problem of predicting the streamwise $uu$ and spanwise $ww$ turbulence intensities at the hub height of single wind turbine. We consider a 2D computational domain of size $L_x\times L_z = 5 \times 4$ where $x\in[\,0,~\, 5\,]$ and $z\in[\,-2,~\, 2\,]$. The turbine of unit diameter is located at $x = 2$ and $z = 0$. We use $N_x=13$ and $N_z=9$ equally spaced collocation points to discretize the computational domain rendering to the state in model~\eqref{eq.LTI-model-2} $\bv \in \bbR^{270 \times 1}$. We use LES generated turbulent intensities at various locations within the computational domain to train our stochastic dynamical models. For consistency, we also use all data points before the turbine to match the inflow turbulence conditions with that of LES. We consider three cases in which the available training dataset contains 3 streams of streamwise and spanwise turbulent intensity measurements from behind the blade tips (edges of 2D rotor structure) and the turbine nacelle (middle of rotor structure) and at various distances away from the turbine: (i) at the turbine location $x=2$ and points within one diameter away (\textbf{Figure~\ref{fig.WF1x1_uuNM_1D}} and~\textbf{\ref{fig.WF1x1_wwNM_1D}}), (ii) at $x=2$ and points within 2 diameters away (\textbf{Figures~\ref{fig.WF1x1_uuNM_2D}} and~\textbf{\ref{fig.WF1x1_wwNM_2D}}), and (iii) at $x=2$ and points within 3 diameters away (\textbf{Figures~\ref{fig.WF1x1_uuNM_3D}} and~\textbf{\ref{fig.WF1x1_wwNM_3D}}). As evident from \textbf{Figures~\ref{fig.WF1x1_uuNM_3D}} and~\textbf{\ref{fig.WF1x1_wwNM_3D}}, for the considered turbine and atmospheric conditions, access to flow statistics $3$ diameters away from the turbine can significantly improve the completion of the statistical signature of the flow at hub height. 
Our results demonstrate the ability of the data-enhanced linearized NS equations in capturing the dominant trends of $uu$ and $ww$ in the wake of a turbine.

\begin{figure}
\centering
\begin{tabular}{cccc}
	\hspace{-.8cm} 
	\subfigure[]{\label{fig.SingleTurbine}}
	&&
	\hspace{.4cm}
	\subfigure[]{\label{fig.Umean-singleturbine}}
	&
	\\[-.6cm]
	&
	\begin{tabular}{c}
	\includegraphics[height=.15\textwidth]{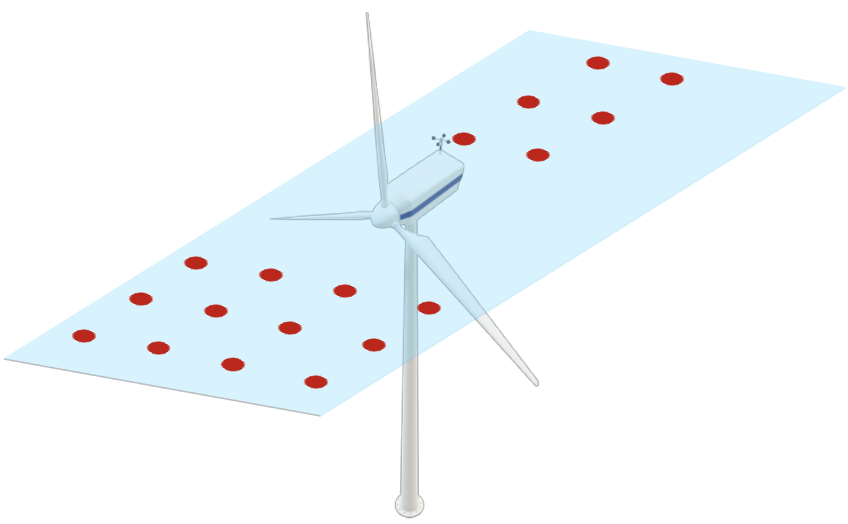}
	\end{tabular}
	&&
	\vspace{.15cm}
    \centering
    \begin{tabular}{cc}
    \hspace{-0.6cm}
        \begin{tabular}{c}
        		\vspace{2cm}
        		\rotatebox{90}{$z$}
        \end{tabular}
    &
    \hspace{-.5cm}
    \includegraphics[width=0.5\textwidth]{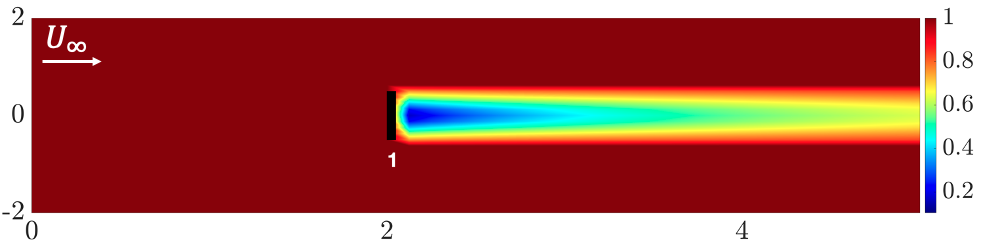}
    \\[-1.2cm]
    &
    \hspace{-.6cm}
    $x$
    \end{tabular}
    \end{tabular}
    \vspace{-.2cm}
	\caption{(a) Schematic of hub-height computational plane with data points used for training in Section~\ref{sec.single-turbine} highlighted in red; (b) Hub-height streamwise velocity $\bar{\bu}(x,z)$ generated using the analytical wake-expansion model of Bastankhah and Port{\'e}-Agel~\cite{baspor14} around which we linearize the NS equations.}
	\vspace{-.2cm}
\end{figure}

\subsubsection{Predicting wind farm turbulence impinging on a cascade of turbines}

\begin{figure}
    \hspace{-0.6cm}
    \begin{tabular}{cccc}
         \subfigure[]{\label{fig.WF4x1_uuLES}}
         &  
         &
         \subfigure[]{\label{fig.WF4x1_wwLES}}
         &
         \\
         \begin{tabular}{c}
		    \vspace{.5cm}
		    \rotatebox{90}{$z$}
	    \end{tabular}
         & 
         \hspace{-0.5cm}
         \begin{tabular}{c}
            \includegraphics[width = 0.45\textwidth]{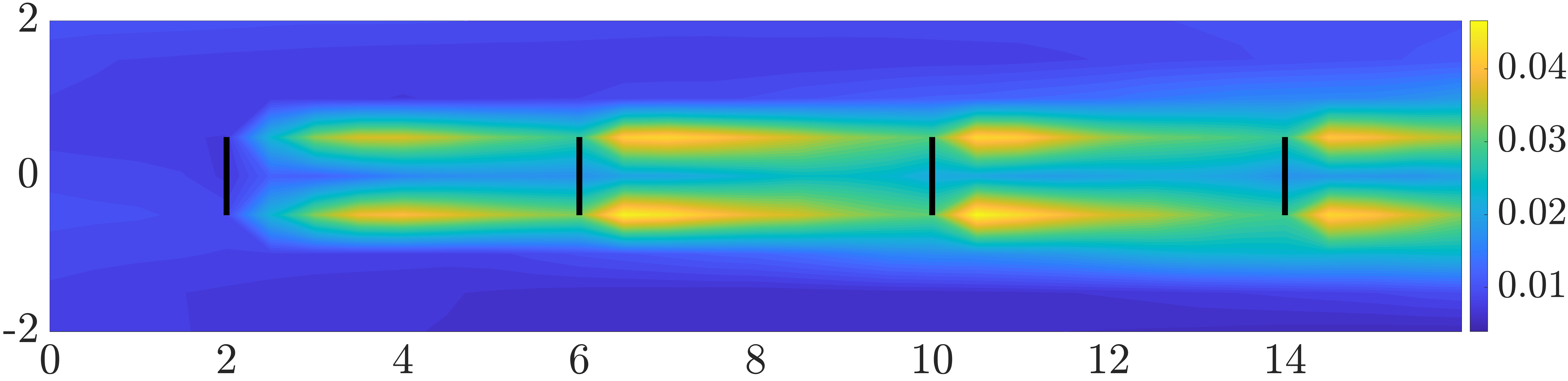}
         \end{tabular}
         &
         & 
         \begin{tabular}{c}
            \includegraphics[width = 0.45\textwidth]{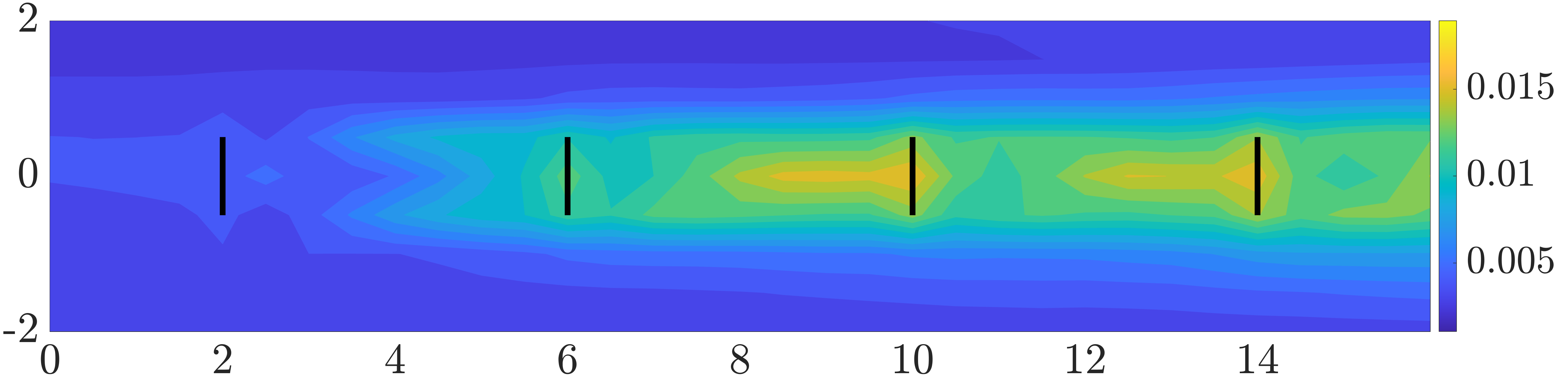}
         \end{tabular}
         \\
         \subfigure[]{\label{fig.WF4x1_uuNM}}
         &  
         &
         \subfigure[]{\label{fig.WF4x1_wwNM}}
         &
         \\
         \begin{tabular}{c}
		    \vspace{.5cm}
		    \rotatebox{90}{$z$}
	    \end{tabular}
         & 
         \hspace{-0.5cm}
         \begin{tabular}{c}
            \includegraphics[width = 0.45\textwidth]{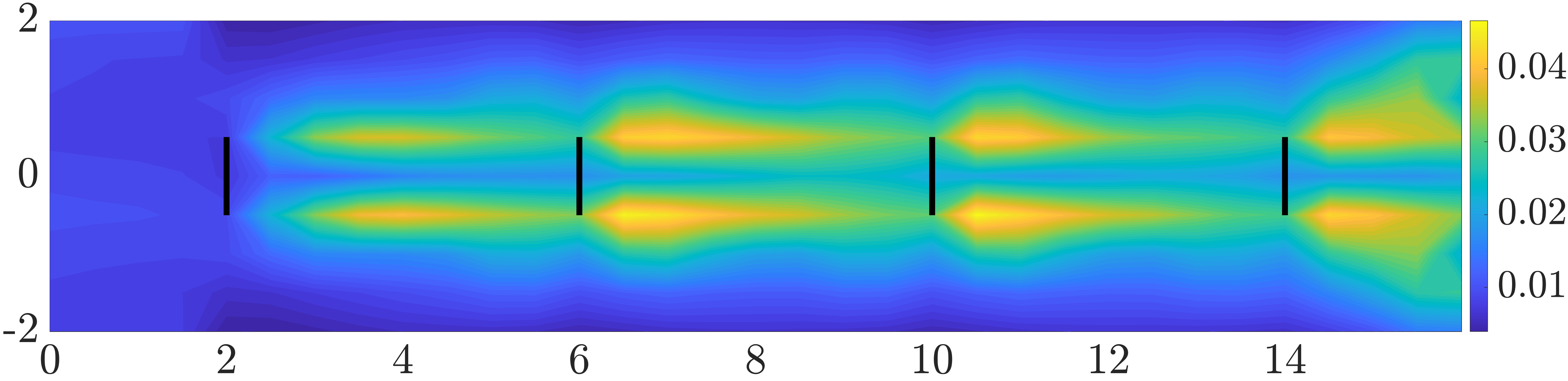}
         \end{tabular}
         &
         & 
         \begin{tabular}{c}
            \includegraphics[width = 0.45\textwidth]{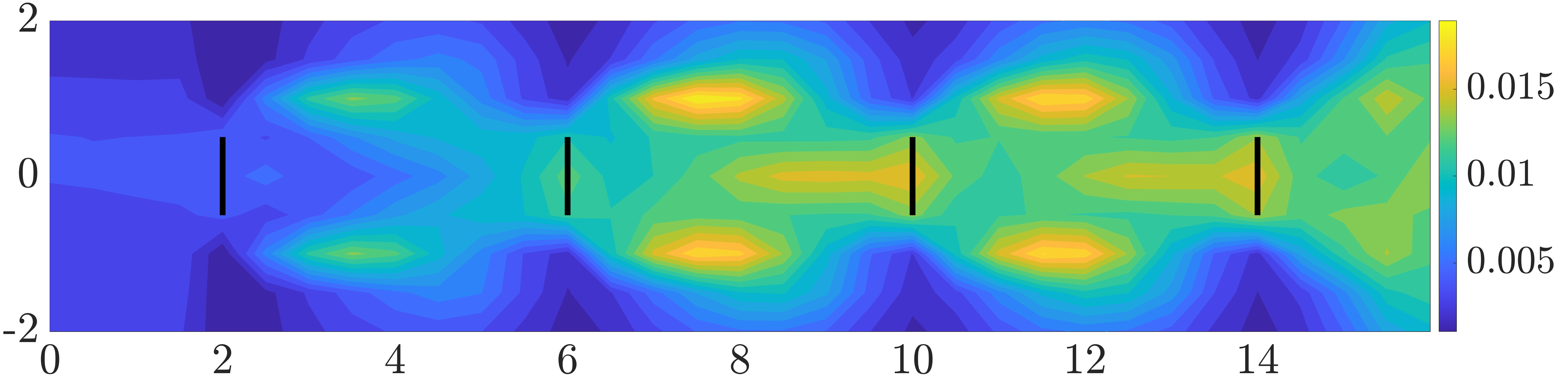}
         \end{tabular}
         \\
         \subfigure[]{\label{fig.Gmatrixuu}}
         &  
         &
         \subfigure[]{\label{fig.Gmatrixww}}
         &
         \\[-.2cm]
         \begin{tabular}{c}
		    \vspace{.5cm}
		    \rotatebox{90}{$z$}
	    \end{tabular}
         & 
         \hspace{-0.5cm}
         \begin{tabular}{c}
            \includegraphics[width = 0.45\textwidth]{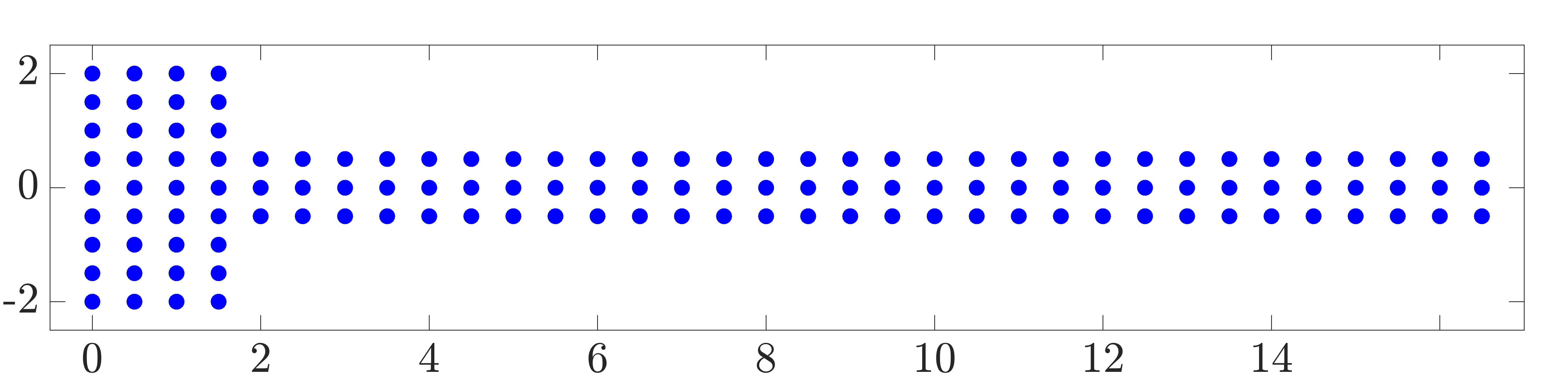}
            \\
            $x$
         \end{tabular}
         &
         & 
         \begin{tabular}{c}
            \includegraphics[width = 0.45\textwidth]{Gmatrix.png}
            \\
            $x$
         \end{tabular}
    \end{tabular}
    \vspace{-.2cm}
    \caption{Streamwise $uu$ (left) and spanwise $ww$ (right) turbulence intensities resulting from LES (a,b) and our stochastic dynamical models (c,d) trained using all intermediate locations downstream of the turbine nacelle and blade tips shown by the blue dots in the plots on the last row.}
    \label{fig.uuww4X1}
\end{figure}

We further extend our study to the case of a $4\times 1$ cascade of turbines that are aligned with the (streamwise) direction of the wind. We consider a similar 2D computational domain of size $L_x\times L_z = 16 \times 9$ where $x\in[\,0,~\, 16\,]$ and $z\in[\,-2,~\, 2\,]$. Turbines of unit diameter are located at $x= \{2, 6, 10, 14\}$ and $z = 0$. We use $N_x=38$ and $N_z=9$ equally spaced collocation points to discretize the computational domain rendering to the state in model~\eqref{eq.LTI-model-2} $\bv \in \bbR^{684 \times 1}$. Given the findings of the single-turbine experiment, we use streamwise and spanwise intensities within $3$ diameters behind the tips and nacelle of each of the turbines as data to train or stochastic models of the hub-height velocity field. \textbf{Figure~\ref{fig.uuww4X1}} demonstrates the performance of our data-enhanced stochastic model in predicting turbulence intensities at hub height. While the overall energy of the flow has been over-predicted by our model (as evident from the energetic patches throughout the farm), dominant features of the streamwise velocity correlations, including regions of high and low energy, are particularly well captured. The stochastic model is also shown to capture the spanwise asymmetry of flow intensities with respect to the centerline running through the turbine nacelles, which is attributed to the turbine's rotation as captured by the high-fidelity LES (\textbf{Figure~\ref{fig.WF4x1_uuNM}}). \textbf{Figure~\ref{fig.WF4x1_wwNM}} shows that the high spanwise intensity regions behind the turbine nacelles are also captured very well by our models albeit spurious regions of high intensity appear in regions slightly beyond the the blade tips. Uncovering potential reasons behind such irregular predictions in the spanwise intensity calls for additional in-depth examination.
Nevertheless, the good quality of completion shown in \textbf{Figure~\ref{fig.uuww4X1}} demonstrates the ability of our linear stochastic dynamical models in predicting the dominant statistical features of the flow and is attributed to the Lyapunov-like constraint in covariance completion problem~\eqref{eq.CC}~\cite{zarjovgeoJFM17,zargeojovARC20}, which keeps physics in the mix and enforces consistency between data and the linearized NS dynamics. 

\vsp

In optimization problem~\ref{eq.CC}, the regularization parameter $\gamma$ determines the importance of the nuclear norm of matrix $Z$ relative to the logarithmic barrier function of the covariance matrix $X$. Larger values of $\gamma$ yield lower-rank matrices $Z$, but may compromise the quality of completion; see~\cite[Appendix C]{zarjovgeoJFM17}. In this study, $\gamma=100$ was observed to provide the best quality of reproduction of the turbulence intensities of velocity fluctuations for both the single- and multi-turbine case studies. In training the stochastic model for the $4\times 1$ cascade of turbines (\textbf{Figure~\ref{fig.uuww4X1}}), where $N_x=38$ and $N_z=9$ ($Z$ is a square matrix of size $684$), $\gamma=100$ results in the $Z$ matrix that solves problem~\eqref{eq.CC} having a rank of $266$ with $265$ positive and $1$ negative eigenvalues. As discussed in Section~\eqref{sec.stat-constraints}, the presence of both positive and negative eigenvalues in matrix $Z$ indicates that the second-order statistics of wind farm turbulence cannot be reproduced by the linearized NS equations with white-in-time stochastic excitation. The distribution of eigenvalues of matrix $Z$ also indicates that $265$ colored-in-time inputs are required to reproduce the partially available entries in covariance matrix $X$ corresponding to the known velocity intensities; see Zare et al.~\cite{zarchejovgeoTAC17,zarjovgeoJFM17} for additional details.

\subsection{Verification in stochastic linear simulations}

\begin{figure}
    \centering
    \begin{tabular}{cc}
         \begin{tabular}{c}
            \vspace{5cm}
            \rotatebox{90}{Kinetic energy}
         \end{tabular}
         &
         \hspace{-.3cm}
            \includegraphics[width = 0.36\textwidth]{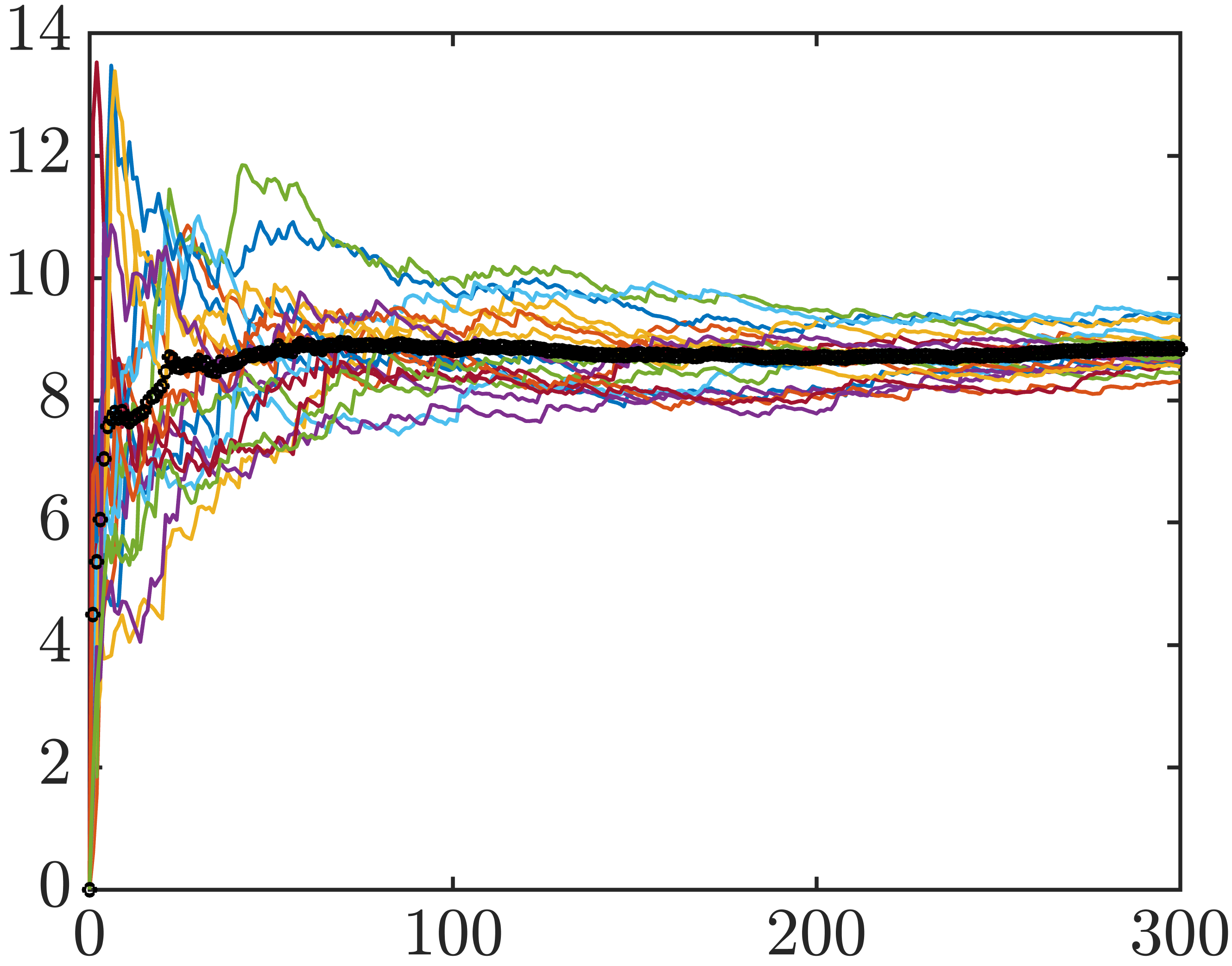} 
            \\[-3.2cm]
            &
            {time} 
    \end{tabular}
    \vspace{-.2cm}
    \caption{Time evolution of fluctuation kinetic energy for 20 realizations of the forcing to the modified linearized dynamics; the energy averaged over all simulations is marked by the thick black line.}
    \label{fig.StochasticSimulations}
\end{figure}

As discussed in Section~\ref{sec.filter-design}, $Z$ can be decomposed into $B H^* + H B^*$ with the input matrix $B$ having 265 independent columns. In other words, the identified X can be explained by driving the LTI model~\eqref{eq.LTI-model-2} with 265 stochastic inputs $d$. The solution to the covariance completion problem~\eqref{eq.CC} also determines the dynamics of the linear filter~\eqref{eq.filter} that generates the coloured-in-time forcing $d$ with appropriate power spectral density. We conduct  stochastic linear simulations to verify our stochastic model of wind farm turbulence (Equation~\eqref{eq.LTI-model-2}). Since a proper comparison with LES requires ensemble averaging, rather than comparison at the level of individual stochastic simulations, we have conducted 20 simulations of system~\eqref{eq.LTI-model-2}). The total simulation time was set to $300$ time units. \textbf{Figure~\ref{fig.StochasticSimulations}} shows the time evolution of the energy (variance) of velocity fluctuations for 20 realizations of white-in-time forcing of the filter dynamics generating the colored-in-time input $\bd$ and exciting the linear dynamical model~\eqref{eq.LTI-model-2}. The variance averaged over all simulations is marked by the thick black line, which asymptotically approaches the value of the total turbulent kinetic energy (averaged over space) in statistical steady state, $\trace(X)$. For the above simulations the final average value has a $4.1\%$ error in matching the training data provided by LES. This close agreement can be further improved by running additional linear simulations and by increasing the total simulation times.

\section{Concluding Remarks}
\label{sec.remarks}

We provide a framework for the stochastic dynamical modeling of wind farm turbulence with enhanced predictive capability relative to conventional low-fidelity models that provide a static (albeit analytical) description of turbine wakes. We focus on the estimation of quantities that are pertinent to control, i.e., thrust forces, power generation, and turbulence intensities throughout wind farms. To capture the complex dynamical nature of wake turbulence,
our proposed approach uses experimentally or numerically generated wind farm quantities to train data-enhanced physics-based models of flow fluctuations that match the available data and complete the statistical signature of the flow. We leverage the predictive capability of the linearized NS equations subject to judiciously shaped additive stochastic excitation as a physics-based model that can overcome potential robustness issues of solely data-driven models in the presence of highly variable atmospheric turbulence. The low-complexity and dynamic nature of this class of models is particularly desirable in handling varying atmospheric conditions that may necessitate online parametric updates based on SCADA measurements. The characteristic features of our models that render them desirable for estimation and control are their: (i) physics-based dynamic nature; (ii) linearity; (iii) low computational complexity; and (iv) statistical consistency in matching flow quantities that are of interest in flow analysis and control.

\vsp

In this paper, we utilize time-averaged LES-generated measurements of thrust forces and/or power generation in addition to turbulence intensities to identify stochastic realizations of forcing into linear approximations of the turbulent flow dynamics to achieve consistency in matching statistical quantities of interest. To demonstrate the utility of our approach, we use the stochastically forced linearized NS equations around a 2D static velocity profile of a wind farm consisting of a cascade of $4$ turbine and show that stochastic modeling of input forcing allows us to significantly improve the predictions of low-fidelity analytical models. We provide details on how matching thrust force (power generation) measurements across various turbines can lead to improved predictions of power generation (thrust force). We also demonstrate the value of turbulent intensity measurements at various distances behind wind turbines in completing the statistical signature of hub-height turbulence, which include dominant features of the velocity correlations and regions of high and low turbulence. We verify our stochastic dynamical models using inexpensive stochastic linear simulations that also highlight the ease of using our low-complexity models for generating statistically consistent turbulent inflow conditions for numerical simulations.

\vsp

We emphasize that the proposed framework allows for linearization around more complicated (potentially 3D) base flow profiles that can better represent the effects of turbine yawing (e.g., wake curl and deflection) or alternative turbine arrangements within a wind farm. Our ongoing efforts involve the development of 3D extensions of the model that resolve the velocity field down to surface and can enable ground sensing capabilities in wind farms, the use such models for sequential data-assimilation, e.g., Kalman filtering, with applications to real-time wind forecasting, and the use of alternative covariance completion formulations~\cite{zarmohdhigeojovTAC20} that may provide useful information about critical directions that have maximal effect in bringing model (in our case the stochastically forced linearized NS) and statistics in agreement. Given the physics-based nature of our models, the latter research direction can prove critical in identifying salient dynamical couplings and interactions in turbine wakes thereby opening the door to new classes of low-fidelity wake models.





\subsection*{Financial disclosure}

None reported.

\subsection*{Conflict of interest}

The authors declare no potential conflict of interests.





\appendix

\section{System matrices in linearized NS equations in evolution form and boundary conditions} 
\label{sec.appendixA}

The system matrices in Equation~\eqref{eq.LTI-model-2} are given as
\begin{align*}
    A &\;=\; 
    \Delta^{-1}
    \tbt{A_{11}}{A_{12}}{A_{21}}{A_{22}},
    \quad\quad
    B \;=\; \Delta^{-1}
    \tbt{{\bf f}_{zz} + 2{\bf f}_z\;\partial_z + {\bf f}\;\partial_{zz}}
    {-({\bf f}_{xz} + {\bf f}_x\;\partial_z + {\bf f}_z\;\partial_x + {\bf f}\;\partial_{xz})}
    {-({\bf f}_{xz} + {\bf f}_x\;\partial_z + {\bf f}_z\;\partial_x + {\bf f}\;\partial_{xz})}
    {{\bf f}_{xx} + 2{\bf f}_x\;\partial_x + {\bf f}\;\partial_{xx}},
    \\[.1cm]
    A_{11} 
    &\;=\;
    -\bar{\bu}\,\Delta\,\partial_x \,-\, \bar{\bu}_x\, \Delta \,-\, 2\, \bar{\bu}_{xz}\, \partial_z \,-\,  \bar{\bu}_{zz}\partial_x \,-\, \bar{\bu}_{xzz} \,+\, \dfrac{1}{Re}\,\Delta^2,
    \\[0.1cm]
    A_{12} 
    &\;=\;
    -\bar{\bu}_{zzz} \,+\, \bar{\bu}_z\, \Delta \,+\, \bar{\bu}_{xz}\,\partial_x \,-\, 2\,\bar{\bu}_{zz}\,\partial_z,
    \\[0.1cm]
    A_{21} 
    &\;=\;
    2\,\bar{\bu}_x\, \partial_{xz} \,+\, \bar{\bu}_{xz}\,\partial_x \,+\, \bar{\bu}_{xxz} \,+\, \bar{\bu}_{xx}\,\partial_z,
    \\[0.1cm]
    A_{22} 
    &\;=\;
    -\bar{\bu}_{xx}\;\partial_{x} \,-\, \bar{\bu}_x\, \Delta\, \partial_x \,-\, 2 \bar{\bu}_{x}\, \partial_{xx} \,+\, \bar{\bu}_{zz}\,\partial_x \,-\, \bar{\bu}_{xzz} \,+\, \dfrac{1}{Re}\,\Delta^2.
\end{align*}
where, ${\bf f}(x,z)$ in matrix $B$ is a 2D shape function that determines the spatial extent of the forcing. For discretization of the domain and finite-dimensional approximation of the differential operators in the system matrices above, we use a second-order central differencing scheme with $N_x$ and $N_z$ uniformly distributed collocation points in the streamwise and spanwise directions, respectively. At the lateral edges of the computational domain, we enforce homogeneous Dirichlet and Neumann boundary conditions, i.e., $\bv(x,z(1)) = \bv(x,z(N_z)) = \bv_x(x,z(1)) = \bv_x(x,z(N_z)) = \bv_z(x,z(1)) = \bv_z(x,z(N_z)) = 0$. At the inlet and outlet of the domain along the streamwise dimension, we apply linear extrapolation conditions (see~\cite{ranzarhacjovPRF19b} for details), i.e., 
\begin{align*}
        \bv(x(1),z) \;=\; \alpha\,\bv(x(2),z) \,+\, \beta\,\bv(x(3),z),\quad\quad
        &
        \bv(x(N_x),z) \;=\; \alpha\,\bv(x(N_x-1),z) \,+\, \beta\,\bv(x(N_x-2),z),
        \\[.15cm]
        \bv_x(x(1),z) \;=\; \alpha\,\bv_x(x(2),z) \,+\, \beta\,\bv_x(x(3),z),\quad\quad
        &  
        \bv_x(x(N_x),z) \;=\; \alpha\,\bv_x(x(N_x-1),z) \,+\, \beta\,\bv_x(x(N_x-2),z),
        \\[.15cm]
        \bv_z(x(1),z) \;=\; \alpha\,\bv_z(x(2),z) \,+\, \beta\,\bv_z(x(3),z),\quad\quad
        &  
        \bv_z(x(N_x),z) \;=\; \alpha\,\bv_z(x(N_x-1),z) \,+\, \beta\,\bv_z(x(N_x-2),z)
        \\[.25cm]
        \alpha \;=\; \dfrac{x(N_x) - x(N_x-2)}{x(N_x-1) - x(N_x-2)},\quad\quad
        &
        \beta \;=\; \dfrac{x(N_x-1) - x(N_x)}{x(N_x-1) - x(N_x-2)}.
\end{align*}
Note that in the case of an equally spaced grid, $\alpha=2$ and $\beta=-1$. We also introduce sponge layers at the inflow and outflow to mitigate the influence of boundary conditions on the fluctuation dynamics within the computational domain~\cite{niclel11,man12,ranzarnicjovAIAA17}. 

\section{Training data} 
\label{sec.appendixB}

Given a time-averaged thrust force measurement $\bar{F}_i$ for the $i$th turbine segment, the effective flow intensity at the staggered point corresponding to that segment follows from equation~\eqref{eq.F-expansion} as
\begin{align}
    \label{eq.vf}
    \overline{\bv^2}_{F,i}  \;=\; \bar{F}_i/\Big(\dfrac{1}{2}\, \rho\, A_i\, C_T \Big) \,-\, \bar{\bu}_{\mathrm{eff},i}^2.
\end{align}
On the other hand, if instead we are provided with a time averaged power measurement $\bar{P}_i$, the effective flow intensity at the staggered point corresponding to that segment is given by
\begin{align}
    \label{eq.vp}
    \overline{\bv^2}_{P,i} \;=\; \left(\bar{P}_i/\Big(\dfrac{1}{2}\, \rho\, A_i\, C_P\, \Big)\right)^{2/3} -\, \bar{\bu}_{\mathrm{eff},i}^2.
\end{align}
Therefore, if thrust forces are provided, we may model the stochastic velocity field $\bv$ to match the effective intensity $\overline{\bv^2}_{F,i}$ and predict $\bar{P}_i$. Similarly, if power measurements are provided, we may model $\bv$ to match the effective intensity $\overline{\bv^2}_{P,i}$ in~\eqref{eq.vp} and predict $\bar{F}_i$. However, due to a lack of sufficient degrees of freedom in Equations~\eqref{eq.FP-expansions}, both thrust force and power measurements cannot be simultaneously matched. To provide a balanced approximation of both, the velocity field $\bv$ can be modeled to match a balanced intensity $\overline{\bv^2}_{\mathrm{bal}}$ per rotor segment as the solution to the problem:
\begin{align}
	\minimize\limits_{\overline{\bv^2}_{\mathrm{bal},i} 
	\,\geq\, 0}
	\quad
    	w_F\,|\overline{\bv^2}_{\mathrm{bal},i} \,-\, \overline{\bv^2}_{F,i}|
    	\;+\; 
    	w_P\, |\overline{\bv^2}_{\mathrm{bal},i} \,-\, \overline{\bv^2}_{P,i}|
	\label{eq.least-squares}
\end{align}
where weights $w_F$ and $w_P$ may be empirically determined to signify the importance of measurements over different turbine segments.


\begin{thebibliography}{100}
\providecommand \doibase [0]{http://dx.doi.org/}%

\bibitem{fleannshawananazhahutwancheche17}
Fleming P, Annoni J, Shah JJ, et al. Field test of wake steering at an offshore
  wind farm. {\it Wind Energy Sci.} 2017\string; 2(1)\string: 229-239.

\bibitem{ahmbasahscougirkazmat19}
Ahmad T, Basit A, Ahsan M, et al. Implementation and analyses of yaw based
  coordinated control of wind farms. {\it Energies} 2019\string; 12(7)\string:
  1266.

\bibitem{duccougirgiegoc19}
Duc T, Coupiac O, Girard N, Giebel G, G{\"o}{\c{c}}men T. Local turbulence
  parameterization improves the Jensen wake model and its implementation for
  power optimization of an operating wind farm. {\it Wind Energy Sci.}
  2019\string; 4(2)\string: 287-302.

\bibitem{howleldab19}
Howland MF, Lele KS, Dabiri JO. Wind farm power optimization through wake
  steering. {\it Proc. Natl. Acad. Sci.} 2019\string; 116(29)\string:
  14495-14500.

\bibitem{flekinsimroaschmurlun20}
Fleming P, King J, Simley E, et al. Continued results from a field campaign of
  wake steering applied at a commercial wind farm--Part 2. {\it Wind Energy
  Sci.} 2020\string; 5(3)\string: 945-958.

\bibitem{bosrui20}
Bossanyi EA, Ruisi R. Axial induction controller field test at Sedini wind
  farm. {\it Wind Energy Sci. Discussions} 2020\string; 2020\string: 1-27.

\bibitem{doekermatkan21}
Doekemeijer BM, Kern S, Maturu S, et al. Field experiment for open-loop
  yaw-based wake steering at a commercial onshore wind farm in Italy. {\it Wind
  Energy Sci.} 2021\string; 6(1)\string: 159-176.

\bibitem{simflegirall21}
Simley E, Fleming P, Girard N, Alloin L, Godefroy E, Duc T. Results from a
  wake-steering experiment at a commercial wind plant: investigating the wind
  speed dependence of wake-steering performance. {\it Wind Energy Sci.}
  2021\string; 6(6)\string: 1427-1453.

\bibitem{johfri12}
Johnson KE, Fritsch G. Assessment of extremum seeking control for wind farm
  energy production. {\it Wind Engineering} 2012\string; 36(6)\string: 701-715.

\bibitem{crelisee09}
Creaby J, Li Y, Seem JE. Maximizing wind turbine energy capture using
  multivariable extremum seeking control. {\it Wind Engineering} 2009\string;
  33(4)\string: 361-387.

\bibitem{cirrotleo17}
Ciri U, Rotea MA, Leonardi S. Model-free control of wind farms: A comparative
  study between individual and coordinated extremum seeking. {\it Renewable
  energy} 2017\string; 113\string: 1033-1045.

\bibitem{cirrotsanleo17}
Ciri U, Rotea MA, Santoni C, Leonardi S. Large-eddy simulations with
  extremum-seeking control for individual wind turbine power optimization. {\it
  Wind Energy} 2017\string; 20(9)\string: 1617-1634.

\bibitem{cirleorot19}
Ciri U, Leonardi S, Rotea MA. Evaluation of log-of-power extremum seeking
  control for wind turbines using large eddy simulations. {\it Wind Energy}
  2019\string; 22(7)\string: 992-1002.

\bibitem{kumrot22}
Kumar D, Rotea MA. {Wind Turbine Power Maximization Using Log-Power
  Proportional-Integral Extremum Seeking}. {\it Energies} 2022\string;
  15(3)\string: 1004.

\bibitem{zhodoy98}
Zho K, Doyle JC. {\it Essentials of robust control}. 104.
\newblock Prentice hall Upper Saddle River, NJ .
\newblock 1998.

\bibitem{skopos07}
Skogestad S, Postlethwaite I. {\it Multivariable feedback control: analysis and
  design}. 2.
\newblock Wiley New York .
\newblock 2007.

\bibitem{goimey15}
Goit JP, Meyers J. Optimal control of energy extraction in wind-farm boundary
  layers. {\it J. Fluid Mech.} 2015\string; 768\string: 5-50.

\bibitem{goimunmey16}
Goit JP, Munters W, Meyers J. Optimal coordinated control of power extraction
  in LES of a wind farm with entrance effects. {\it Energies} 2016\string;
  9(1)\string: 29.

\bibitem{munmey17}
Munters W, Meyers J. An optimal control framework for dynamic induction control
  of wind farms and their interaction with the atmospheric boundary layer. {\it
  Philosophical Transactions of the Royal Society A: Mathematical, Physical and
  Engineering Sciences} 2017\string; 375(2091)\string: 20160100.

\bibitem{munmey18}
Munters W, Meyers J. Towards practical dynamic induction control of wind farms:
  analysis of optimally controlled wind-farm boundary layers and sinusoidal
  induction control of first-row turbines. {\it Wind Energy Sci.} 2018\string;
  3(1)\string: 409-425.

\bibitem{munmey18ENG}
Munters W, Meyers J. Dynamic strategies for yaw and induction control of wind
  farms based on large-eddy simulation and optimization. {\it Energies}
  2018\string; 11(1)\string: 177.

\bibitem{doevanfle19}
Doekemeijer BM, {van Wingerden} JW, Fleming PA. A tutorial on the synthesis and
  validation of a closed-loop wind farm controller using a steady-state
  surrogate model. {\it 2019 American Control Conference (ACC)}. Philadelphia, PA, USA: IEEE; 2019\string: 2825-2836.

\bibitem{sinsei19}
Singh P, Seiler P. Controlling the meandering wake using measurement feedback. {\it 2019 American Control Conference (ACC)}. Philadelphia, PA, USA: IEEE; 2019\string: 4144-4150.

\bibitem{doevamvan20}
Doekemeijer BM, {van der Hoek} D, {van Wingerden} JW. Closed-loop model-based
  wind farm control using {FLORIS} under time-varying inflow conditions. {\it
  Renew. Energy} 2020\string; 156\string: 719-730.

\bibitem{jen83}
Jensen NO. A note on wind generator interaction. {\it Roskilde, Denmark:
  Ris{\o} National Laboratory} 1983.

\bibitem{katic86}
Katic I, H{\o}jstrup J, Jensen NO. A simple model for cluster efficiency. {\it 1986 European wind energy association conference and exhibition, vol. 1}. Rome, Italy; 1986\string: 407-410.

\bibitem{ain88}
Ainslie JF. Calculating the flowfield in the wake of wind turbines. {\it J.
  Wind Eng. Ind. Aerodyn.} 1988\string; 27(1-3)\string: 213-224.

\bibitem{burjenshabos11}
Burton T, Jenkins N, Sharpe D, Bossanyi E. {\it Wind energy handbook}.
\newblock John Wiley \& Sons .
\newblock 2011.

\bibitem{frandsen06}
Frandsen S, Barthelmie R, Pryor S, et al. Analytical modelling of wind speed
  deficit in large offshore wind farms. {\it Wind Energy: An International
  Journal for Progress and Applications in Wind Power Conversion Technology}
  2006\string; 9(1-2)\string: 39-53.

\bibitem{baspor14}
Bastankhah M, Port{\'e}-Agel F. A new analytical model for wind-turbine wakes.
  {\it Renewable energy} 2014\string; 70\string: 116-123.

\bibitem{wantanchogu16}
Wang L, Tan AC, Cholette M, Gu Y. Comparison of the effectiveness of analytical
  wake models for wind farm with constant and variable hub heights. {\it Energy
  Conversion and Management} 2016\string; 124\string: 189-202.

\bibitem{cammolschbot19}
Campagnolo F, Molder A, Schreiber J, Bottasso C. Comparison of analytical wake
  models with wind tunnel data. {\it J. Phys.: Conf. Ser.}, 1256. IOP Publishing; 2019\string:
  012006.

\bibitem{zhaletiun20}
Zhan L, Letizia S, Iungo G. Optimal tuning of engineering wake models through
  lidar measurements. {\it Wind Energy Sci.} 2020\string; 5(4)\string:
  1601-1622.

\bibitem{marannflechu19}
Mart{\'\i}nez-Tossas LA, Annoni J, Fleming PA, Churchfield MJ. The aerodynamics
  of the curled wake: a simplified model in view of flow control. {\it Wind
  Energy Sci.} 2019\string; 4(1)\string: 127-138.

\bibitem{marbra20}
Mart{\'\i}nez-Tossas LA, Branlard E. The curled wake model: equivalence of shed
  vorticity models. {\it J. Phys.: Conf. Ser.}, 1452. IOP Publishing; 2020\string: 012069.

\bibitem{zonpor20}
Zong H, Port{\'e}-Agel F. A point vortex transportation model for yawed wind
  turbine wakes. {\it J. Fluid Mech} 2020\string; 890.

\bibitem{basshashagaymen22}
Bastankhah M, Shapiro CR, Shamsoddin S, Gayme DF, Meneveau C. A vortex sheet
  based analytical model of the curled wake behind yawed wind turbines. {\it J.
  Fluid Mech.} 2022\string; 933.

\bibitem{larmadtholar08}
Larsen GC, Madsen HA, Thomsen K, Larsen TJ. Wake meandering: a pragmatic
  approach. {\it Wind Energy} 2008\string; 11(4)\string: 377-395.

\bibitem{annhowseigua16}
Annoni J, Howard K, Seiler P, Guala M. An experimental investigation on the
  effect of individual turbine control on wind farm dynamics. {\it Wind Energy}
  2016\string; 19(8)\string: 1453-1467.

\bibitem{guorotsum20}
Guo Y, Rotea M, Summers T. Stochastic dynamic programming for wind farm power
  maximization. {\it 2019 American Control Conference (ACC)}: IEEE; 2020\string: 4824-4829.

\bibitem{iunviocirleorot16}
Iungo GV, Viola F, Ciri U, Leonardi S, Rotea M. Reduced order model for
  optimization of power production from a wind farm. 34th Wind Energy Symposium; 2016\string: 2200.

\bibitem{boedoevalmeyvan18}
Boersma S, Doekemeijer B, Vali M, Meyers J, Wingerden vJW. A control-oriented
  dynamic wind farm model: {WFSim}. {\it Wind Energy Sci.} 2018\string;
  3(1)\string: 75-95.

\bibitem{letiun22}
Letizia S, Iungo GV. {Pseudo-2D RANS: A LiDAR-driven mid-fidelity model for
  simulations of wind farm flows}. {\it J. Renew. Sustain. Energy} 2022\string;
  14(2)\string: 023301.

\bibitem{annbayjohdalquokemfle19}
Annoni J, Bay C, Johnson K, et al. Wind direction estimation using {SCADA} data
  with consensus-based optimization. {\it Wind Energy Sci.} 2019\string;
  4(2)\string: 355-368.

\bibitem{bercirrotleo21}
Bernardoni F, Ciri U, Rotea MA, Leonardi S. Identification of wind turbine
  clusters for effective real time yaw control optimization. {\it J. Renew.
  Sustain. Energy} 2021\string; 13(4)\string: 043301.

\bibitem{stastamengaykin21}
Starke GM, Stanfel P, Meneveau C, Gayme DF, King J. Network based estimation of
  wind farm power and velocity data under changing wind direction. {\it 2021 American Control Conference (ACC)}: IEEE;
  2021\string: 1803-1810.

\bibitem{vermen14}
VerHulst C, Meneveau C. Large eddy simulation study of the kinetic energy
  entrainment by energetic turbulent flow structures in large wind farms. {\it
  Phys. Fluids} 2014\string; 26\string: 025113.

\bibitem{anngebsei16}
J.~Annoni PG, Seiler P. Wind farm flow modeling using an input-output
  reduced-order model. {\it 2016 American Control Conference (ACC)}: IEEE; 2016\string: 506-512.

\bibitem{raaschche17}
S.~Raach DS, Cheng PW. Lidar-based wake tracking for closed-loop wind farm
  control. {\it Wind Energy Sci.} 2017\string; 2(1)\string: 257-267.

\bibitem{sinpaojen20}
Sinner M, Pao LY, King J. Estimation of large-scale wind field characteristics
  using supervisory control and data acquisition measurements. {\it 2020 American Control Conference (ACC)}: IEEE;
  2020\string: 2357-2362.

\bibitem{noamortad11}
Noack BR, Morzy{\'n}ski M, Tadmor G. {\it Reduced-order modelling for flow
  control}. 528 of {\it {CISM} Courses and Lectures}.
\newblock Springer .
\newblock 2011.

\bibitem{tadnoa11}
Tadmor G, Noack BR. Bernoulli, {B}ode, and {B}udgie [{A}sk the {E}xperts]. {\it
  IEEE Contr. Syst. Mag.} 2011\string; 31(2)\string: 18-23.

\bibitem{wanwuxia17}
Wang JX, Wu JL, Xiao H. Physics-informed machine learning approach for
  reconstructing {R}eynolds stress modeling discrepancies based on {DNS} data.
  {\it Phys. Rev. Fluids} 2017\string; 2(3)\string: 034603.

\bibitem{wuxiapat18}
Wu JL, Xiao H, Paterson E. Physics-informed machine learning approach for
  augmenting turbulence models: {A} comprehensive framework. {\it Phys. Rev.
  Fluids} 2018\string; 3(7)\string: 074602.

\bibitem{karkevluperrwanyan21}
Karniadakis GE, Kevrekidis IG, Lu L, Perdikaris P, Wang S, Yang L.
  Physics-informed machine learning. {\it Nat. Rev. Phys.} 2021\string;
  3(6)\string: 422-440.

\bibitem{solwisbra14}
Soleimanzadeh M, Wisniewski R, Brand A. State-space representation of the wind
  flow model in wind farms. {\it Wind Energy} 2014\string; 17(4)\string:
  627-639.

\bibitem{boevalkuhvan16}
Boersma S, Vali M, K{\"u}hn M, {van Wingerden} JW. Quasi linear parameter
  varying modeling for wind farm control using the {2D Navier-Stokes}
  equations. {\it 2020 American Control Conference (ACC)}: IEEE; 2016\string: 4409-4414.

\bibitem{hoechebewhen05}
H{\oe}pffner J, Chevalier M, Bewley TR, Henningson DS. State estimation in
  wall-bounded flow systems. {P}art 1. {P}erturbed laminar flows. {\it J.\
  Fluid Mech.} 2005\string; 534\string: 263-294.

\bibitem{chehoebewhen06}
Chevalier M, H{\oe}pffner J, Bewley TR, Henningson DS. State estimation in
  wall-bounded flow systems. {P}art 2. {T}urbulent flows. {\it J.\ Fluid Mech.}
  2006\string; 552\string: 167-187.

\bibitem{butfar92}
Butler KM, Farrell BF. Three-Dimensional Optimal Perturbations in Viscous Shear
  Flow. {\it Phys.\ Fluids A} 1992\string; 4\string: 1637.

\bibitem{tretrereddri93}
Trefethen LN, Trefethen AE, Reddy SC, Driscoll TA. Hydrodynamic Stability
  Without Eigenvalues. {\it Science} 1993\string; 261\string: 578-584.

\bibitem{farioa93}
Farrell BF, Ioannou PJ. Stochastic Forcing of the Linearized {N}avier-{S}tokes
  Equations. {\it Phys.\ Fluids A} 1993\string; 5(11)\string: 2600-2609.

\bibitem{bamdah01}
Bamieh B, Dahleh M. Energy Amplification in Channel Flows with Stochastic
  Excitation. {\it Phys.\ Fluids} 2001\string; 13(11)\string: 3258-3269.

\bibitem{jovbamJFM05}
Jovanovic MR, Bamieh B. Componentwise energy amplification in channel flows.
  {\it J. Fluid Mech.} 2005\string; 534\string: 145-183.

\bibitem{ranzarhacjovPRF19b}
Ran W, Zare A, Hack MJP, Jovanovic MR. Stochastic receptivity analysis of
  boundary layer flow. {\it Phys. Rev. Fluids} 2019\string; 4(9)\string: 093901
  (28 pages).

\bibitem{mcksha10}
McKeon BJ, Sharma AS. A critical-layer framework for turbulent pipe flow. {\it
  J.\ Fluid Mech.} 2010\string; 658\string: 336-382.

\bibitem{hwacosJFM10b}
Hwang Y, Cossu C. Linear non-normal energy amplification of harmonic and
  stochastic forcing in the turbulent channel flow. {\it J.\ Fluid Mech.}
  2010\string; 664\string: 51-73.

\bibitem{zarjovgeoJFM17}
Zare A, Jovanovic MR, Georgiou TT. Colour of turbulence. {\it J. Fluid Mech.}
  2017\string; 812\string: 636-680.

\bibitem{deljim03}
{Del {\'A}lamo} JC, Jim{\'e}nez J. Spectra of the very large anisotropic scales
  in turbulent channels. {\it Phys.\ Fluids} 2003\string; 15(6)\string: 41-44.

\bibitem{deljimzanmos04}
{Del {\'A}lamo} JC, Jim{\'e}nez J, Zandonade P, Moser RD. Scaling of the energy
  spectra of turbulent channels. {\it J.\ Fluid Mech.} 2004\string;
  500(1)\string: 135-144.

\bibitem{siljimmos13}
Sillero JA, Jim{\'e}nez J, Moser RD. One-point statistics for turbulent
  wall-bounded flows at {Reynolds} numbers up to $\delta^+ \approx 2000$. {\it
  Phys. Fluids} 2013\string; 25\string: 105102.

\bibitem{siljimmos14}
Sillero JA, Jim{\'e}nez J, Moser RD. Two-point statistics for turbulent
  boundary layers and channels at {Reynolds} numbers up to $\delta^+ \approx
  2000$. {\it Phys. Fluids} 2014\string; 26\string: 105109.

\bibitem{hutmonganngmar11}
Hutchins N, Monty JP, Ganapathisubramani B, Ng HCH, Marusic I.
  Three-dimensional conditional structure of a high-{R}eynolds-number turbulent
  boundary layer. {\it J. Fluid Mech.} 2011\string; 673\string: 255-285.

\bibitem{schfla13}
Schultz MP, Flack KA. Reynolds-number scaling of turbulent channel flow. {\it
  Phys. Fluids} 2013\string; 25\string: 025104.

\bibitem{zarchejovgeoTAC17}
Zare A, Chen Y, Jovanovic MR, Georgiou TT. Low-complexity modeling of partially
  available second-order statistics: theory and an efficient matrix completion
  algorithm. {\it IEEE Trans. Automat. Control} 2017\string; 62(3)\string:
  1368-1383.

\bibitem{zarjovgeoCDC16}
Zare A, Jovanovic MR, Georgiou TT. Perturbation of system dynamics and the
  covariance completion problem. {\it 2016 IEEE 55th Conference on Decision and Control (CDC)}: IEEE; 2016\string: 7036-7041.

\bibitem{zargeojovARC20}
Zare A, Georgiou TT, Jovanovic MR. Stochastic dynamical modeling of turbulent
  flows. {\it Annu. Rev. Control Robot. Auton. Syst.} 2020\string; 3\string:
  195-219.

\bibitem{zarmohdhigeojovTAC20}
Zare A, Mohammadi H, Dhingra NK, Georgiou TT, Jovanovi\'c MR. Proximal
  algorithms for large-scale statistical modeling and sensor/actuator
  selection. {\it IEEE Trans. Automat. Control} 2020\string; 65(8)\string:
  3441-3456.

\bibitem{porluwu10}
Port{\'e}-Agel F, Lu H, Wu YT. A large-eddy simulation framework for wind
  energy applications. {\it The fifth international symposium on computational wind engineering}; 23. Chapel Hill, NC, USA; 2010\string: 27.

\bibitem{wupor11}
Wu YT, Port{\'e}-Agel F. Large-eddy simulation of wind-turbine wakes:
  evaluation of turbine parametrisations. {\it Bound.-Layer Meteorol.}
  2011\string; 138(3)\string: 345-366.

\bibitem{sch10}
Schmid PJ. Dynamic mode decomposition of numerical and experimental data. {\it
  J.\ Fluid Mech.} 2010\string; 656\string: 5-28.

\bibitem{jovschnicPOF14}
Jovanovic MR, Schmid PJ, Nichols JW. Sparsity-promoting dynamic mode
  decomposition. {\it Phys. Fluids} 2014\string; 26(2)\string: 024103 (22
  pages).

\bibitem{annnicsei16}
Annoni JR, Nichols J, Seiler PJ. Wind farm modeling and control using dynamic
  mode decomposition. {\it 34th Wind Energy Symposium}; 2016\string: 2201.

\bibitem{sch22}
Schmid PJ. Dynamic mode decomposition and its variants. {\it Annu. Rev. Fluid
  Mech.} 2022\string; 54\string: 225-254.

\bibitem{khaangparcal00}
Khadra K, Angot P, Parneix S, Caltagirone J. Fictitious domain approach for
  numerical modelling of {N}avier-{S}tokes equations. {\it Int. J. Numer.
  Methods Fluids} 2000\string; 34(8)\string: 651-684.

\bibitem{geo02a}
Georgiou TT. The structure of state covariances and its relation to the power
  spectrum of the input. {\it IEEE Trans. Autom. Control} 2002\string;
  47(7)\string: 1056-1066.

\bibitem{geo02b}
Georgiou TT. Spectral analysis based on the state covariance: the maximum
  entropy spectrum and linear fractional parametrization. {\it IEEE Trans.
  Autom. Control} 2002\string; 47(11)\string: 1811-1823.

\bibitem{faz02}
Fazel M. {\it Matrix rank minimization with applications}. PhD thesis. Stanford
  University; 2002.

\bibitem{recfazpar10}
Recht B, Fazel M, Parrilo PA. Guaranteed minimum-rank solutions of linear
  matrix equations via nuclear norm minimization. {\it SIAM Rev.} 2010\string;
  52(3)\string: 471-501.

\bibitem{SDPT3}
Toh KC, Todd MJ, T{\"u}t{\"u}nc{\"u} RH. {SDPT3 -- a MATLAB software package
  for semidefinite programming, version 1.3}. {\it Optimization methods and
  software} 1999\string; 11(1-4)\string: 545-581.

\bibitem{cvx}
Grant M, Boyd S. {CVX}: {M}atlab {S}oftware for {D}isciplined {C}onvex
  {P}rogramming, version 2.1. \url{http://cvxr.com/cvx};  2014.

\bibitem{boyvan04}
Boyd S, Vandenberghe L. {\it Convex optimization}.
\newblock Cambridge University Press .
\newblock 2004.

\bibitem{zarjovgeoACC15}
Zare A, Jovanovic MR, Georgiou TT. Alternating direction optimization
  algorithms for covariance completion problems. {\it 2015 American Control Conference (ACC)}: IEEE; 2015; Chicago,
  IL\string: 515-520.

\bibitem{NREL-5MW}
Jonkman J, Butterfiled S, Musial W, Scott G. Definition of a 5-MW {R}eference
  {W}ind {T}urbine for {O}ffshore {S}ystem {D}evelopment. Tech. Rep.
  NREL/TP-500-38060, NREL - National Renewable Energy Laboratory; Golden, CO,
  USA:   2009.

\bibitem{santoni2015}
Santoni C, Ciri U, Rotea M, Leonardi S. Development of a high fidelity CFD code
  for wind farm control. {\it 2015 American Control Conference (ACC)}: IEEE; 2015\string: 1715-1720.

\bibitem{santoni2017}
Santoni C, Carrasquillo K, Arenas-Navarro I, Leonardi S. Effect of tower and
  nacelle on the flow past a wind turbine. {\it Wind Energy} 2017\string;
  20(12)\string: 1927-1939.

\bibitem{Ciri17}
Ciri U, Petrolo G, Salvetti MV, Leonardi S. Large-eddy simulations of two
  in-line turbines in a wind tunnel with different inflow conditions. {\it
  Energies} 2017\string; 10(6)\string: 821.

\bibitem{ciri2018}
Ciri U, Salvetti M, Carrasquillo K, Santoni C, Iungo G, Leonardi S. Effects of
  the subgrid-scale modeling in the large-eddy simulations of wind turbines.
  {\it Direct and large-eddy simulation x}: Springer.  2018 (pp. 109-115).

\bibitem{Orlandi06}
Orlandi P, Leonardi S. DNS of turbulent channel flows with two- and
  three-dimensional roughness. {\it J. Turbul.} 2006\string; 7\string: 1-22.

\bibitem{orlanski1976}
Orlanski I. A simple boundary condition for unbounded hyperbolic flows. {\it
  Journal of computational physics} 1976\string; 21(3)\string: 251-269.

\bibitem{johnson2006}
Johnson KE, Pao LY, Balas MJ, Fingersh LJ. Control of variable-speed wind
  turbines: standard and adaptive techniques for maximizing energy capture.
  {\it IEEE Control Systems Magazine} 2006\string; 26(3)\string: 70-81.

\bibitem{laks2009}
Laks JH, Pao LY, Wright AD. Control of wind turbines: Past, present, and
  future. {\it 2009 American Control Conference}: IEEE; 2009\string: 2096-2103.

\bibitem{leonardi2010}
Leonardi S, Castro IP. Channel flow over large cube roughness: a direct
  numerical simulation study. {\it Journal of Fluid Mechanics} 2010\string;
  651\string: 519-539.

\bibitem{jonbutmussco09}
Jonkman J, Butterfield S, Musia W, Scott G. Definition of a {5-MW} reference
  wind turbine for offshore system development. tech. rep., National Renewable
  Energy Lab. (NREL); Golden, CO, USA:   2009.

\bibitem{santcarareleo17}
Santoni C, Carrasquillo K, Arenas-Navarro I, Leonardi S. Effect of tower and
  nacelle on the flow past a wind turbine. {\it Wind Energy} 2017\string;
  20(12)\string: 1927-1939.

\bibitem{sangarcirzhaiunleo20}
Santoni C, Garc{\'\i}a-Cartagena EJ, Ciri U, Zhan L, Iungo GV, Leonardi S.
  {One-way mesoscale-microscale coupling for simulating a wind farm in North
  Texas: Assessment against SCADA and LiDAR data}. {\it Wind Energy}
  2020\string; 23(3)\string: 691-710.

\bibitem{iunwupor13}
Iungo GV, Wu YT, Port{\'e}-Agel F. Field measurements of wind turbine wakes
  with lidars. {\it J. Atmos. Ocean. Technol.} 2013\string; 30(2)\string:
  274-287.

\bibitem{niclel11}
Nichols JW, Lele SK. Global modes and transient response of a cold supersonic
  jet. {\it J. Fluid Mech.} 2011\string; 669\string: 225-241.

\bibitem{man12}
Mani A. Analysis and optimization of numerical sponge layers as a nonreflective
  boundary treatment. {\it J. Comput. Phys.} 2012\string; 231(2)\string:
  704-716.

\bibitem{ranzarnicjovAIAA17}
Ran W, Zare A, Nichols JW, Jovanovic MR. The effect of sponge layers on global
  stability analysis of Blasius boundary layer flow. 47th AIAA Fluid Dynamics Conference. Denver, CO, USA; 2017\string: 3456 (12 pages).

\end{thebibliography}
\end{document}